\documentclass[%
 reprint,
 amsmath,amssymb,
 aps,
]{revtex4-2}
\usepackage{xcolor}
\usepackage{graphicx}
\usepackage{caption}
\usepackage{subcaption}
\usepackage{comment}
\usepackage{multirow}
\usepackage{amsmath,mathtools}
\DeclareMathOperator{\polylog}{polylog}
\usepackage[hidelinks]{hyperref}
\hypersetup{
    colorlinks,
    linkcolor={red!70!black},
    citecolor={blue},
    urlcolor={blue!70!black}
}

\newcommand \maxcut {MaxCut }
\newcommand \sm {Supplementary Material}
\newcommand \mcp {\maxcut problem }
\newcommand \bluenum {B }
\newcommand \qemc {QEMC }
\newcommand{\subf}[2]{%
  {\small\begin{tabular}[t]{@{}c@{}}
  #1\\#2
  \end{tabular}}%
}

\newcommand{\ket}[1]{\left|#1\right\rangle }

\newcommand{\ceil}[1]{\left\lceil #1 \right\rceil}

\usepackage{dcolumn}
\usepackage{bm}

\begin{document}
\preprint{APS/123-QED}

\title{A Variational Qubit-Efficient MaxCut Heuristic Algorithm}

\author{Yovav Tene-Cohen}
\thanks{Equal contribution.}
\affiliation{The Engineering Faculty, Bar-Ilan University, Ramat-Gan 52900, Israel}
\affiliation{Center for Quantum Entanglement Science and Technology, Bar-Ilan University, 52900 Ramat-Gan, Israel}

\author{Tomer Kelman}
\thanks{Equal contribution.}
\affiliation{The Engineering Faculty, Bar-Ilan University, Ramat-Gan 52900, Israel}
\affiliation{Center for Quantum Entanglement Science and Technology, Bar-Ilan University, 52900 Ramat-Gan, Israel}

\author{Ohad Lev}
\affiliation{The Engineering Faculty, Bar-Ilan University, Ramat-Gan 52900, Israel}
\affiliation{Center for Quantum Entanglement Science and Technology, Bar-Ilan University, 52900 Ramat-Gan, Israel}

\author{Adi Makmal}
\affiliation{The Engineering Faculty, Bar-Ilan University, Ramat-Gan 52900, Israel}
\affiliation{Center for Quantum Entanglement Science and Technology, Bar-Ilan University, 52900 Ramat-Gan, Israel}

\date{\today}

\begin{abstract}
MaxCut is a key NP-Hard combinatorial optimization graph problem with extensive theoretical and industrial applications, including the Ising model and chip design. 
While quantum computing offers new solutions for such combinatorial challenges which are potentially better than classical schemes, with the Quantum Approximate Optimization Algorithm (QAOA) being a state-of-the-art example, its performance is currently hindered by hardware noise and limited qubit number. 
Here, we present a new variational Qubit-Efficient MaxCut (QEMC) algorithm that requires a logarithmic number of qubits with respect to the graph size, an exponential reduction compared to QAOA. We demonstrate cutting-edge performance for graph instances consisting of up to 32 nodes (5 qubits) on real superconducting hardware, and for graphs with up to 2048 nodes (11 qubits) using noiseless simulations, outperforming the established classical algorithm of Goemans and Williamson (GW). The QEMC algorithm's innovative encoding scheme empowers it with great noise-resiliency on the one hand, but also enables its efficient classical simulation on the other, thus obscuring a distinct quantum advantage. Nevertheless, even in the absence of quantum advantage, the QEMC algorithm serves as a potential quantum-inspired algorithm, provides a challenging benchmark for QAOA, and presents a novel encoding paradigm with potential applications extending to other quantum and classical algorithms.
\end{abstract}

\maketitle

\section{\label{sec:intro} Introduction}
Quantum computing offers promising prospects for achieving highly efficient computation schemes that surpass the capabilities of standard classical approaches  \cite{2011_Nielsen_Xhuang_QI}. Yet, realizing useful quantum computations on the current noisy intermediate scale quantum (NISQ) technology is a worldwide challenge \cite{2018_Quantum_Preskill_NISQ}.  

Variational quantum algorithms (VQAs) are a family of hybrid quantum-classical algorithms which provides a heuristic approach for solving different types of optimization problems via an iterative tuning of parameterized quantum circuits. It includes the variational quantum eigensolver (VQE) for finding ground-states of molecular Hamiltonians \cite{peruzzo2014variational, tilly2022variational}, quantum machine-learning algorithms (see \cite{mangini2023variational} and references therein), and more \cite{bravo2020variational}.
The quantum approximate optimization algorithm (QAOA) \cite{farhi2014quantum_org_qaoa} is a VQA that is used to find approximated solutions for combinatorial optimization problems. It promises an optimal solution in the limit of infinite number of quantum circuit layers and was studied extensively in the context of the \maxcut problem \cite{crooks2018performance_QAOA, 2020_PRX_Lukin_QAOA_performance, wurtz2021fixed, harrigan2021quantum,weidenfeller2022scaling}, defined below. 
The QAOA utilizes a basis encoding, meaning that the computational basis state of each qubit ($\ket{0}$ or $\ket{1}$) encodes the logical state of the corresponding problem variable. 
This implies that the number of required qubits for QAOA scales linearly with the problem size. In particular, it requires $N$ qubits for solving the \mcp of an $N$-node graph, which limits the usage of the QAOA in current quantum computers
to very small problem instances \cite{Guerrero2020SolvingCO,harrigan2021quantum, weidenfeller2022scaling, guerreschi2019qaoa}.

\begin{figure*}[ht!]
    \centering    \includegraphics[width=0.75\textwidth]{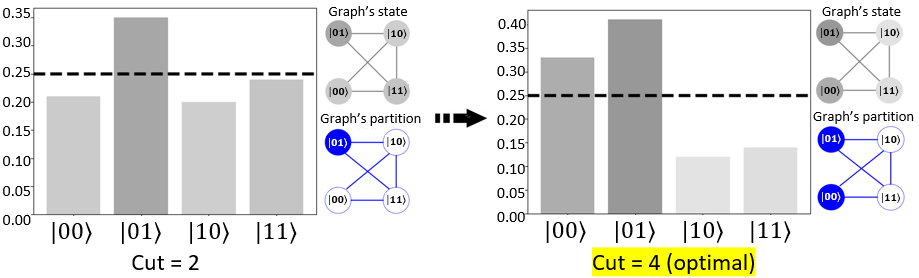}
    \caption{Illustration of the QEMC encoding scheme and optimization process for an $N=4$ nodes graph. \textbf{(a)} \textbf{Left}: An initial random quantum state's probability distribution, along with the corresponding graph's state and logical state (partition), resulting in a cut of 2. \textbf{(b)} \textbf{Right:} A final quantum state's probability distribution,  with the corresponding graph's state and logical state (partition), resulting in a cut of 4. The probability-threshold is determined by setting $B=\frac{N}{2}=2$ blue nodes and is given by $\frac{1}{4}$, as dictated by Eq.~\ref{eq:encoding_rule}.}.
    \label{fig:method_illustration}
\end{figure*}

Recently, several attempts have been made towards devising qubit-efficient VQAs for combinatorial problems. 
To that end, several studies have explored the incorporation of divide-and-conquer techniques to break down the problem into smaller sub-problems, each requiring only a small number of qubits  \cite{dunjko2018computational,ge2020hybrid,guerreschi2021solving,li2022large, zhou2022qaoa_in_qaoa}. Another approach, more closely related to our study, develops and utilizes alternative information encoding schemes that result in a reduction in the number of qubits required. For example, in Ref.~\cite{glos2022space} the traveling salesman problem (TSP) over $N$ cities was encoded using $N \log{N}$ qubits, instead of $N^2$ qubits in the standard encoding. A different study proposed to divide combinatorial problems to $2^{n_r}$ independent subsystems, each with $n_a$ variables, which can be encoded using $n_a+n_r$ qubits \cite{tan2021qubit}. This approach relies on the topology of the problem and lies between two extremes: representing the problem precisely with $N$ qubits for $N$ variables and using $\log{N}$ qubits but neglecting all interrelations between the variables. 
Qubit-efficient encoding for solving the $k$-coloring graph problem \cite{tabi2020quantum} and the Max $k$-Cut problem (a generalization of the MaxCut problem to $k$ partitions) \cite{fuchs2021efficient} were also devised, using $N\log{k}$ qubits instead of $Nk$ qubits. 
In \cite{patti2022variational}, the required number of qubits for solving the MaxCut problem of graphs with $N$ nodes was reduced to $\frac{N}{2}$ qubits using a multibasis graph encoding, which outperformed the QAOA with a comparable circuit Ansatz.
Lastly, a quantum method utilizing $\log{N}$ qubits was introduced in \cite{ranvcic2023noisy,chatterjee2023solving} for solving the MaxCut problem of graphs with $N$ nodes using the graph's Laplacian matrix. However, it does not appear to reach high performance. 

This paper introduces the variational qubit-efficient MaxCut (QEMC) algorithm, designed to find heuristic solutions for the MaxCut problem on an $N$-node graph using $\ceil{\log{N}}$ qubits, an exponential reduction compared to QAOA. Unlike the work presented in \cite{ranvcic2023noisy,chatterjee2023solving}, we do not rely on the Laplacian matrix of the graph. Instead, we achieve this significant qubit reduction via an innovative information encoding scheme based on probability-threshold encoding.

The QEMC algorithm operates outside the QAOA framework. Beyond its unique encoding scheme, it does not embed the graph instance directly into the quantum circuit Ansatz. This approach, combined with the algorithm's logarithmic qubit scaling, allows for the use of significantly shallower circuits, which are less susceptible to noise. Furthermore, the QEMC's cost function is phrased as a continuous mean-squared-error estimator, deviating from the conventional VQA technique of expressing cost functions as expectation values of Hamiltonians.

As shown below, the QEMC algorithm does not offer a straightforward quantum advantage. This is not surprising. In general, and without delving into the specific details of the QEMC algorithm, a classical simulation of a $\log{N}$ qubit circuit requires storage and time that scale approximately linearly with $N$, while the mere specification of the problem (representing the graph, reading it in and out, etc.\!, see also \cite{aaronson2015read}) often scales at least linearly with $N$. Since the quantum algorithm must also specify the problem, it too, requires resources that scale linearly with the size of the problem, thus offering no significant advantage, if at all. This is in contrast to VQAs which utilize the standard basis encoding, where this issue does not arise.

The significance of this paper lies in several aspects: first, the presentation of the QEMC algorithm, a new, qubit-efficient, variational solver for the MaxCut problem that exhibits high performance in comparison to both the classical Goemans-Williamson (GW) algorithm in classical simulations and the quantum QAOA algorithm on actual quantum devices, as detailed below; second, the introduction of a novel information encoding for the MaxCut problem, which can be utilized in other quantum and classical algorithms; third, the potential role of the proposed QEMC algorithm as a new benchmark for the QAOA algorithm; and last, its potential usage as a classical quantum-inspired algorithm.

The remainder of the paper is structured as follows:
Sec.~\ref{sec:maxcut} shortly describes the \maxcut problem and the classical Goemans-Williamson (GW) approximation algorithm, providing essential background information. 
In Sec.~\ref{sec:QEMC}, we introduce the QEMC algorithm, delving into its encoding scheme, cost function, and the unconstrained circuit Ansatz it employs.
The algorithm's performance is experimentally demonstrated in Sec.~\ref{sec:results}, where we use classical simulations on graphs ranging from 16 to 2048 nodes, and real quantum hardware calculations using IBMQ devices, on graphs spanning from 4 to 32 nodes. This section also offers a comparison of QEMC's performance against GW and QAOA. Following this, Sec.~\ref{sec:resources} provides an estimation of the computational resources demanded by the QEMC and its classical simulation counterpart. Finally, in Sec.~\ref{sec:conclusions_discussions}, we conclude the paper, discuss its implications, and outline potential directions for future research.

\section{Background: the \maxcut problem}
\label{sec:maxcut}
 The \maxcut problem is an NP-hard  \cite{Karp1972} combinatorial problem  with practical applications in circuit design and statistical physics \cite{AnApplicationofCombinatorialOptimizationtoStatisticalPhysicsandCircuitLayoutDesign}.  
 It is formalized as  follows:
let $G=(V,E)$ be a (weighted) graph consisting of nodes $V$ and (weighted) edges $E$, then finding the \maxcut amounts to partitioning the nodes into two distinct subsets, e.g. ``blue" and ``white" nodes,  such that the total sum (weight) of the edges that connect nodes from different sets is maximized. In unweighted graphs, each edge connecting between a blue and a white node simply adds +1 to the cut score, whereas an edge connecting two nodes of the same color does not contribute to the  score.

The Goemans-Williamson (GW) is the best-known classical approximation algorithm for the MaxCut problem \cite{goemans1995improved_GW}. It is a polynomial time semidefinite programming algorithm, which guarantees a minimal approximation ratio of: 
\begin{equation}
r_A^{GW}  \cong  \frac{\maxcut_{GW}}{\maxcut_{OPT}} =  0.87856, 
\end{equation}
where $\maxcut_{OPT}$ is the optimal cut.  It often serves as a benchmark for testing new MaxCut heuristics, and utilized similarly in this paper.

\section{The qubit-efficient MaxCut (QEMC) Algorithm}
\label{sec:QEMC}
The QEMC algorithm is based on a novel probability-threshold encoding scheme, a suitable cost function, and a parameterized unconstrained quantum circuit, which we describe below.
Here and throughout the paper, we use $N=|V|$ to denote the number of nodes in the graph and $M=|E|$ to indicate the number of edges. 

\subsection{\label{secsec:info_encoding} The probability-threshold encoding}
The QEMC maps each node $k$ into its corresponding computational basis state $\ket{k}$, using the standard decimal to binary mapping. 
More precisely, given a variational quantum state  $\ket{\psi(\theta)}=\sum_{k=0}^{N-1}\alpha_k (\theta) \ket{k}$ of $\log{N}$ qubits, where $\theta$ stands for the circuit's variational parameters, the logical state of node $k$ (being ``blue" or ``white") is encoded onto the probability $p_{\theta}(k)=|\alpha_k(\theta)|^2$ to measure the computational basis state $\ket{k}$:  a low probability indicates w.l.o.g.\ ``white" nodes, whereas a high probability indicates ``blue" nodes.
For a graph with \textbf{$B$} blue nodes, a possible but naive qubit-efficient mapping would associate an exact zero-probability with white nodes and an exact $\frac{1}{\bluenum}$-probability with blue nodes, to preserve normalization.
However, this definitive mapping assigns meaning only to a vary small part of the Hilbert-space, leaving the vast majority of it uninterpreted. 
Instead, to account for \emph{all} possible states continuously, we set an intermediate probability-threshold of $p_{th}=\frac{1}{2\bluenum}$ that distinguishes between white and blue nodes, as follows: 
\begin{equation}
    \label{eq:encoding_rule}
            color(k)= 
    \begin{cases}
        white,& \text{if } p(k)\leq  p_{th}=\frac{1}{2\bluenum} \\
        blue,              & \text{otherwise}
    \end{cases}
\end{equation}
This is illustrated in Fig.~\ref{fig:method_illustration}.

In essence, this procedure maps each graph partition to \emph{a volume} of quantum states, rather than to just a single state, thereby providing the optimization procedure a large set of optimal quantum states to search for. 
We believe that this property of the algorithm plays a central role in its strength and relative noise resilience, as demonstrated below.

The attentive reader would notice that  Eq.~\ref{eq:encoding_rule} assumes prior knowledge of the number of blue nodes $\bluenum$, which, however, cannot be obtained in advance. Yet, this does not pose a problem as we can efficiently scan through all possible values of $\bluenum = 1,...,\lfloor{\frac{N}{2}}\rfloor$
(it is preferable to designate the smaller set as the blue set because it leads to larger probability-thresholds $p_{th}$, which consequently require less shots for statistical accuracy). 
Moreover, in practice, for many cases such as the regular graphs we examined, it is reasonable to set $B=\frac{N}{2}$ which assumes that half of the nodes are blue and half are white.  
It should be noted that the value of $B$ we assign does not rigidly constrain the algorithm; instead, it generates solutions where the number of blue nodes is approximately $B$, due to state normalization.
Finally, we note that this encoding scheme works equally well for all graph sizes, including graphs whose  number of nodes is not a power of two, through (zero) padding over the non-existing nodes.

\subsection{The QEMC cost function}
The QEMC algorithm searches for quantum states that minimize the following continuous cost function: 
 \begin{equation}
    \label{eq:QEMC_cost}
     \sum_{\substack{
    j<k:
    \\
    \{j,k\} \in E
    }}
    \left[\left(d(j,k)-\frac{1}{B}\right)^2 + \left(s(j,k)-\frac{1}{B}\right)^2\right],
\end{equation}
 where 
the sum goes over all pairs of neighboring nodes $j$ and $k$, and where $d(j,k) = |p(j)-p(k)|$ and $s(j,k) = p(j)+p(k)$ are the absolute difference and sum of the corresponding states probabilities. 
The idea is that when both $d(j,k)$ and $s(j,k)$ approach $\frac{1}{B}$ we get that the probability of one of the nodes approaches zero (distinctive ``white") whereas the probability of the other node approaches $\frac{1}{B}$ (distinctive ``blue"), without imposing which is which. Minimizing the cost function of Eq.~\ref{eq:QEMC_cost} drives the quantum circuit to generate quantum states whose probability histogram maximizes the number of neighboring nodes 
 that fulfill this relation, thereby maximizing the cut in the graph.

\subsection{The QEMC quantum circuit}
The QEMC circuit Ansatz does not depend on specific graph-instances. This is in contrast to the QAOA, in which the graph-structure is encoded explicitly into the quantum circuit \cite{farhi2014quantum_org_qaoa}. Instead, the graph is encoded implicitly through the cost function, as described above. Consequently, the QEMC quantum circuit is not constrained to any particular form and should only be expressive enough to approach the optimal states in the Hilbert-space.  
Such a problem-independent Ansatz approach, denoted as hardware-efficient in \cite{kandala2017hardware}, offers significant flexibility in the choice of the Ansatz. Moreover, it can be potentially implemented via pulse optimization schemes, such as outlined in \cite{meitei2021gate, liang2023napa, meirom2022pansatz,de2023pulse}, whose application on problem-dependent Ans\"atze is not straightforward.

Throughout the optimization procedure, the QEMC circuit is executed and measured multiple times to obtain a statistically reliable probability histogram of the quantum state. 
Then, as with all VQAs, the variational parameters of the circuit $\theta$ are updated using a classical optimizer to optimize the cost function until convergence.  
In Fig.~\ref{fig:method_illustration}, the QEMC encoding scheme and the optimization process for a graph with $N=4$ nodes are demonstrated. Initially, a random distribution is shown on the left, representing a partition with $Cut=2$. The QEMC circuit is then iteratively optimized until it converges to the state in the right panel, whose probability distribution encodes an optimal partition, corresponding to $Cut^*=4$.

\section{\label{sec:results} Results}
To demonstrate the performance of the QEMC method we carried out classical noiseless state-vector simulations, as well as quantum real-hardware IBMQ calculations.
We focused on two sets of graphs: small 3-regular graphs with 4-32 nodes (2-5 qubits), and large 9-regular graphs with 16-2048 nodes (4-11 qubits).
To assess the cuts produced by the QEMC algorithm, we compared them to the optimal cuts obtained by exhaustive search (for the smaller 3-regular graphs only) and to GW cuts (for both graph sets). The latter were derived using the public code from \cite{rvg77_maxcut} available on GitHub, which we have thoroughly verified independently.

\subsection{Computational details}
\label{secsec:computational_details}

\begin{figure}[h!]
    \centering    \includegraphics[width=0.48\textwidth]{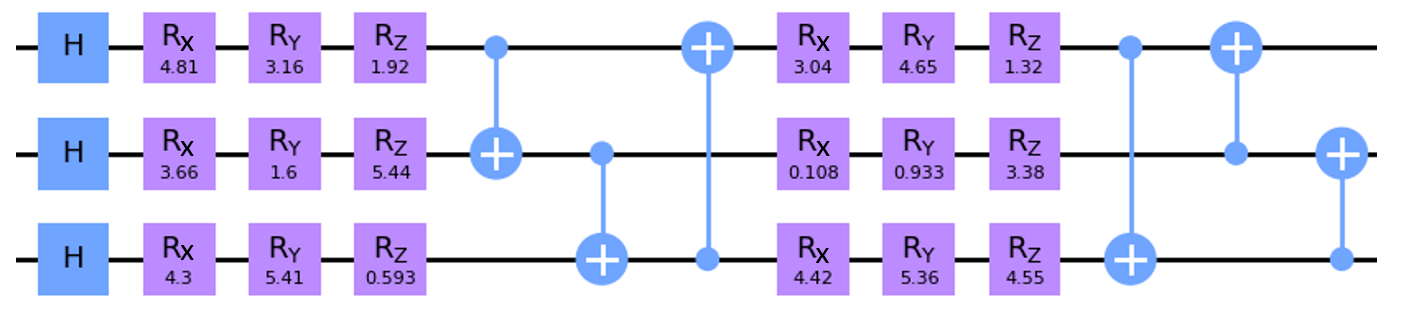}
    \caption{Illustration of PennyLane's ``strongly entangling layers" circuit Ansatz \cite{schuld2020circuit, Pennylane} used in this study. The figure depicts the Ansatz for the case of $n=3$ qubits and $L=2$ layers. After a single (uncounted) layer of Hadamard gates, each subsequent layer consists of $3n$ single-qubit parameterized rotation gates and $n$ CNOT gates.} 
    \label{fig:QEMCSingleLayer}
\end{figure}

Our classical simulations were conducted using the PennyLane platform \cite{Pennylane}. For the real IBMQ executions, we utilized the following machines: 'Jakarta', 'Perth', and 'Lagos', where the specific hardware selection was based on availability. We did not employ any noise-mitigation techniques throughout.

As described above, the QEMC Ansatz is instance-independent and is not restricted to any particular architecture. In this study, we employed  PennyLane's ``strongly entangling layers" circuit \cite{schuld2020circuit,Pennylane}, depicted in Fig.~\ref{fig:QEMCSingleLayer}, in which
the number of tunable parameters in each layer increases linearly with the number of qubits, resulting in logarithmic scaling with the number of nodes in the graph ($N$). 
We did not optimize over this choice. For all calculations, including those conducted on real-hardware, we initialized the quantum circuit with random parameters without optimizing over their initialization (see e.g. \cite{sack2021quantum, amosy2022iterative} for works on circuit parameters initialization for the QAOA).

\begin{figure}[h!]
  \includegraphics[width=0.35\textwidth]{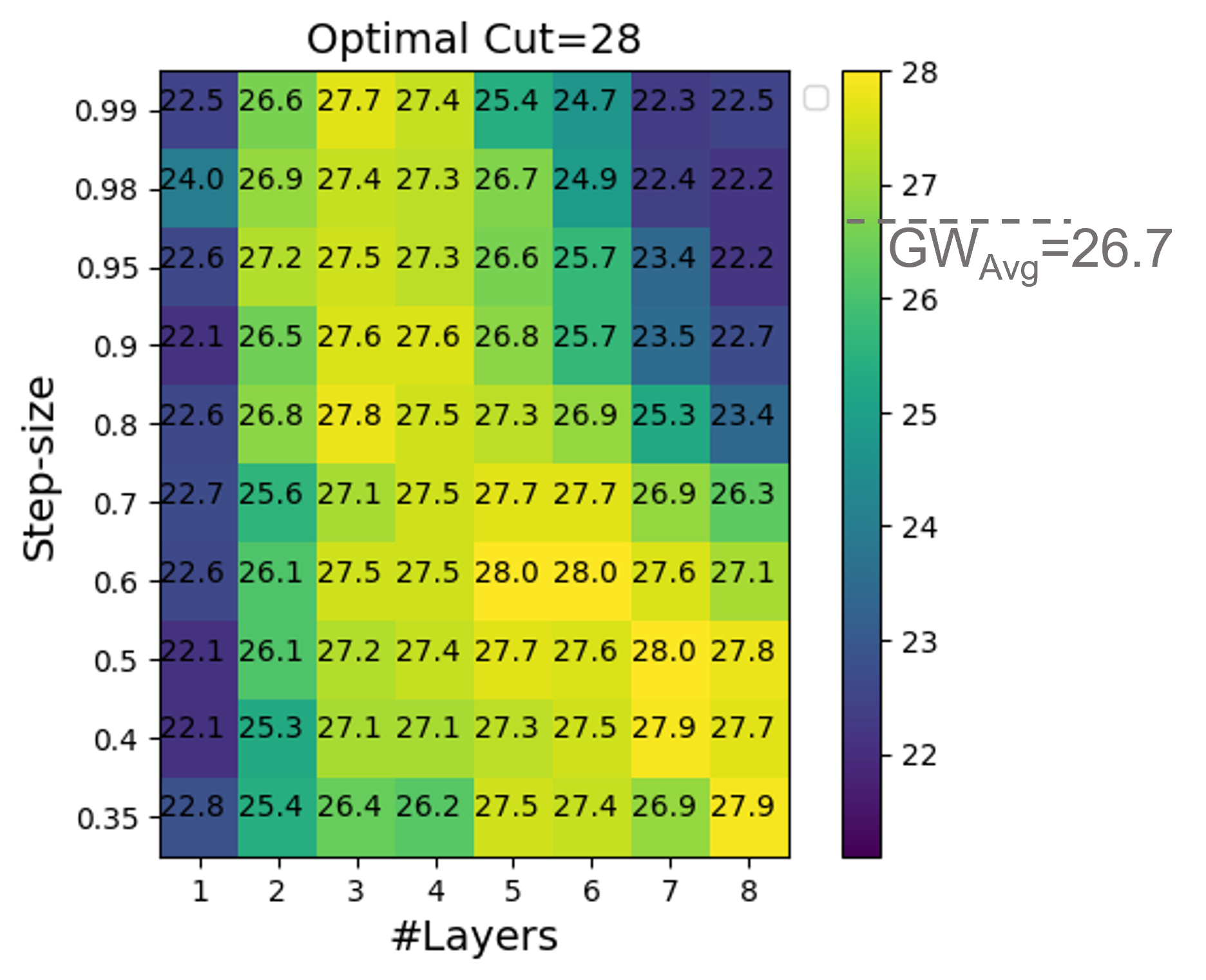}
    \caption{Hyperparameters grid-search on the \textbf{22-node} graph instance which has an optimal cut of 28 and an average GW cut of 26.7. The grid-search is performed to optimize the choice of the step-size in the Adam optimizer and the number of layers in the ``strongly entangling layers" circuit. Each grid entry represents the average QEMC cut obtained from 10 different runs, each with up to 300 iterations, using classical noiseless simulations. The results indicate that the average GW cut is achievable with 2 or more layers, and that finding the optimal cut requires 5 or more layers.
    Additionally, deeper circuits exhibit improved performance with smaller step-sizes, consistently observed across all cases studied.} 
    \label{fig:22_grid_search}
\end{figure}

The Adam optimizer \cite{kingma2014adam} was used with the default hyperparameters of $\beta_1 = 0.9, \beta_2=0.99, \varepsilon=10^{-8}$ fixed throughout, as implemented in PennyLane. We did not optimize over that choice either. 
To fix the $\alpha$ step-size in the Adam optimizer and the number of circuit layers $L$, we ran a grid-search over these two hyperparameters, as depicted in Fig.~\ref{fig:22_grid_search} for a specific graph instance with $22$ nodes. For the noiseless simulations, we chose the minimum number of layers that achieved the largest average cut, together with the corresponding step-size. For example, in the case of the $22$-node graph, presented in Fig.~\ref{fig:22_grid_search}, we set the number of layers to $L = 5$, and the step-size to $\alpha = 0.6$. 
For the real-hardware calculations, we utilized a maximum number of $L = 2$ layers and selected the most suitable corresponding step-size.
By doing that we slightly sacrificed theoretical accuracy, but gained significant practical accuracy, due to usage of shallower circuits. 

We set $B=\frac{N}{2}$ throughout, namely we assumed that optimal MaxCut partitions would have approximately half of the nodes that are blue, and did not optimize over that choice any further.  Finally, we configured the number of shots for the real-hardware calculations to be $S=3N^2$, 
based on an empirical estimation described in Sec.~\ref{sec:resources}.

\subsection{\label{sec:4-32 nodes results} Small 3-regular graphs:  4-32 nodes (2-5 qubits) - noiseless simulations and real-hardware calculations}

\subsubsection{A single graph instance per graph-size}
\label{secsec:4-32-single}

\begin{figure}[h!]
  \begin{subfigure}[]{0.27\textwidth}
\includegraphics[width=\textwidth]{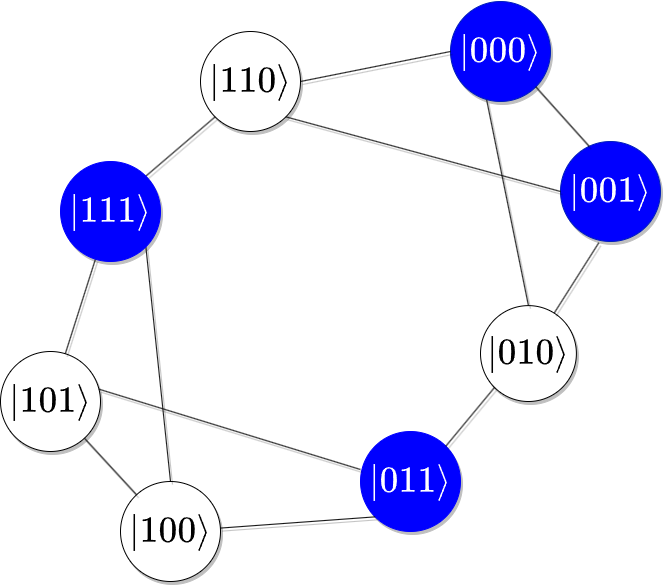}
        \caption{The 8-node graph instance.}
  \label{fig:8_nodes_graph}
  \end{subfigure} 
  \begin{subfigure}[b]{0.38\textwidth}
        \includegraphics[width=\textwidth]{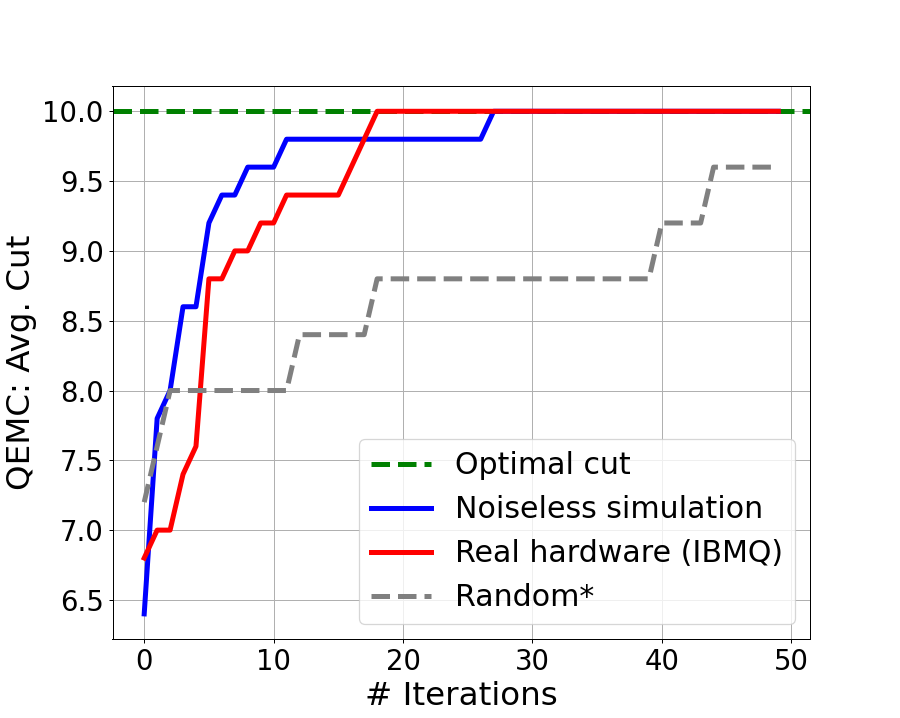}
        \caption{Avg. QEMC best-so-far cut as a function of QEMC-iterations.}
        \label{fig:8_nodes_cuts}
  \end{subfigure}
  \begin{subfigure}[b]{0.38\textwidth}
        \includegraphics[width=\textwidth]{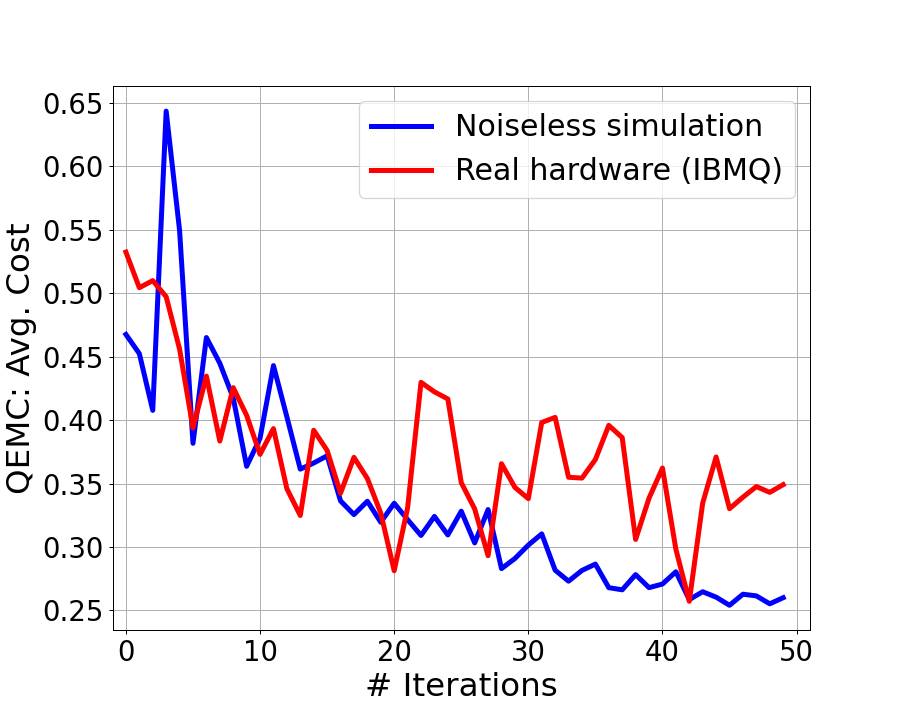}
        \caption{Avg. QEMC cost as a function of QEMC-iterations.}
        \label{fig:8_nodes_cost}
  \end{subfigure}
    \caption{The 8-node graph instance (3 qubits): Noiseless simulations vs. real-hardware (IBMQ) calculations.}
\label{fig:8_nodes}
\end{figure}

We begin with small to medium-sized 3-regular graphs, consisting of 4-32 nodes (increasing in increments of 2), which we encode using 2-5 qubits. For each graph-size, we focus on a single graph instance. 
We utilized the exact graph instances from Ref.~\cite{harrigan2021quantum} for graphs with 4-22 nodes, which form a reference point for our results. 
The remaining graphs, namely 3-regular graphs ranging from 24-32 nodes, were randomly generated using the NetworkX package \cite{networkx}.
The hyperparameters, i.e.\  the step-size $\alpha$ and the number of layers $L$, were fixed as described above in Sec.~\ref{secsec:computational_details} and are detailed in Table.~\ref{table:hyper_parameters}. Fig.~\ref{fig: grid_search_params} in the \sm\ presents the corresponding grid-searches performed for each graph instance.

\begin{table*}[t]
\centering
 \begin{tabular}{||l | l | c c c c c c c c c c c c c c c||} 
 \hline
\multicolumn{2}{||l|}{$\#$ Nodes $(N)$} & 4 & 6 & 8 & 10 & 12 & 14 & 16 & 18 & 20 & 22 & 24 & 26 & 28 & 30 & 32 \\ [0.5ex] 
 \hline\hline
 & step-size   &  0.99 & 0.99 & 0.99   & 0.99  & 0.98 & 0.9 & 0.95 & 0.95 & 0.8 & 0.6 & 0.8 & 0.5 & 0.5 & 0.8 & 0.7 \\
 Noiseless   & $\#$ layers - optimal & 1 & 2  & 2 & 2 & 3 & 3& 5 & 4 & 4 & 5 & 4 & 7 & 7 & 4 & 5\\
  state-vector  & $\#$ layers (to GW) & 1 & 2  & 2 & 2 & 2 & 2& 3 & 3 & 2 & 2 & 3 & 2 & 2 & 3 & 2\\
  simulations & $\#$ iterations (total) &  &   &  &  &  & & $\cdots \quad 300 \quad  \cdots$  &  &  & &  & &  &  & \\
  & $\#$ iterations (to GW) & 1 &  24 & 11 & 20 & 31 & 161& 103 & 81 &  42 & 29 & 54 & 121&  80& 81 & 34\\
 \hline
  & step-size   &  0.9 &  &  0.9 &  & & & 0.7 & & & & & & & & 0.9\\
  & $\#$ layers   &  1 &  &  2 &  & & & 2& & & & & & & & 2 \\
 Real  & $\#$ shots ($3N^2$)  &  48 &  &  192 &  & & &768 & & & & & & & & 2048 \\
-hardware & $\#$ iterations (total)   & 11  &  &  129.2 &  & & & 98.6& & & & & & & & 62.0 \\
(IBMQ) & $\#$ iterations (to GW)    & 3   &  &  17 &  & & & -- & & & & & & & & -- \\
   & Quantum device &  Lagos &  &   Jakarta&  & & & Perth / Jakarta & & & & & & & & Jakarta \\
 \hline
 \hline
   & GW   & 3.7  & 7 & 10  & 11 &15.1&18.6 & 20.3 & 22.8& 24.3& 26.7&30.8 & 32.6& 33.7&37.8 & 38.6\\
  Avg. cut & QEMC simulation  & 4 & 7 & 10 & 12 & 16 & 19 & 20.7 & 23.9 & 26.7 & 28 & 32.6 & 33.3 & 35.6 & 38.9 & 41.5\\
  & QEMC real-hardware & 4  &  & 10  &  & & &18.6 & & & & & & & & 32.0\\
  \hline
   & GW   & 4  & 7 & 10  &12 &16 &19 & 21 &24 &27 &28 &33 &33 &36 &39 & 41\\
  Max cut & QEMC simulation  & 4 & 7 & 10& 12&16 &19 &21 & 24 & 27 & 28 &33 &34 &36 &40 &42 \\
   & QEMC real-hardware & 4  &  &  10 &  & & & 19 & & & & & & & & 34\\
    \hline
  Optimal cut & exhaustive search & 4  & 7  & 10  & 12 &16 &19 &21 & 24& 27&28& 33& 34&36 &40 & 42 \\ 
\hline
\hline
 \end{tabular}
 \caption{\textbf{QEMC hyperparameters and attained cuts for the set of 3-regular graphs, ranging from 4 to 32 nodes, with one graph instance per size.}
$\quad$ \textbf{Noiseless state-vector simulations}: The table specifies the optimal step-size and number of layers, as determined via grid-search, the minimal number of layers required to reach the averaged GW cut, the total number of iterations performed, and the iteration count required to reach the averaged GW cut.
\textbf{Real-hardware IBMQ calculations}: The table details the utilized step-size and number of layers, the actual number of shots, the average total number of iterations (subject to hardware interruptions), and the iteration count to reach the averaged GW cut, when achieved. Specific quantum devices are also listed.
\textbf{Avg. cut:} The highest average cuts of GW, QEMC simulations, and QEMC real-hardware calculations are reported.
\textbf{Max cut:} The maximum cuts from the best trials are reported for GW, QEMC simulations, and QEMC real-hardware calculations.
\textbf{Optimal cut:} Optimal cuts from exhaustive search are specified.}
 \label{table:hyper_parameters}
\end{table*}

\paragraph*{The 8-node graph instance.}
As an elaborated example, we focus on the 8-node graph instance, as depicted  
 in Fig.~\ref{fig:8_nodes_graph}, with an optimal partition yielding $Cut^*=10$,  found through exhaustive search. Within the QEMC algorithm, this graph is encoded using 3 qubits.  
Fig.~\ref{fig:8_nodes_cuts} displays the QEMC best-so-far cut for this graph as a function of QEMC-iterations for both classical noiseless state-vector simulations (blue) and real-hardware IBMQ calculations (red). The green dashed line represents the optimal cut.
The cut values are averaged over 10 executions in the case of classical calculations, and 5 executions for the real-hardware calculations, each with varying random parameter initializations.
As a sanity check, we also plot the best-so-far Random* cut (gray), obtained by randomly sampling graph partitions with $B=\frac{N}{2}=4$ blue nodes, to mimic our partitioning assumption. 
 This sampling method outperforms a naive random sampler in graphs where the optimal partition indeed divides the graph into two equal-sized subsets, as is the case with the graph discussed here, and hence serves as a better and less naive baseline.

It is seen in Fig.~\ref{fig:8_nodes_cuts} that both the noiseless simulations and real-hardware calculations converge rapidly to the optimal cut, with the real-hardware calculations converging even faster. 
Since these results are averaged, we can infer that \emph{all} calculations successfully reached the optimal cut, regardless of the initialization of their circuit parameters. This success of the real-hardware calculations, achieved without any noise mitigation techniques, can be attributed to two factors:  first, the utilization of only 3 qubits and 2 layers; and second, to the threshold probability encoding scheme, 
which allows for a certain degree of variability, thereby enhancing the noise resilience of the scheme. 

For completeness, Fig.~\ref{fig:8_nodes_cost} depicts the average cost function value as a function of QEMC-iterations, as obtained by the noiseless (blue) and real-hardware (red) calculations. It is seen that in both cases the cost decreases rapidly with iterations, as desired. Here, in contrast to the plot in Fig.~\ref{fig:8_nodes_cuts}, the curve does not show monotonic behavior since we plot the actual cost value per QEMC-iteration, rather than the best-so-far one.

\begin{figure}[h!]
    \centering
    \includegraphics[width=0.38\textwidth]{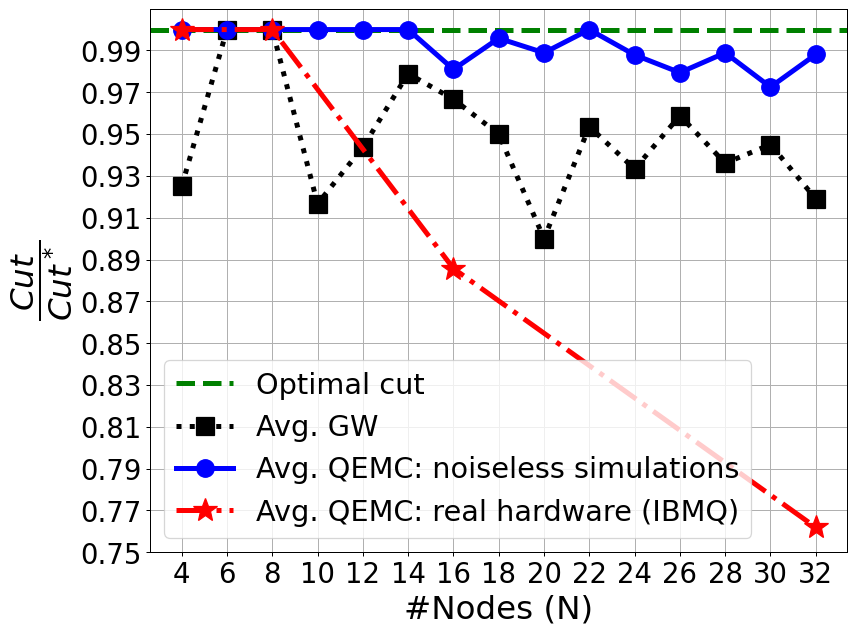}
    \caption{\textbf{Noiseless and real-hardware results for the set of 3-regular graphs with 4 to 32 nodes (2-5 qubits), \textit{one graph instance per size}}. The average cut-ratio, namely the ratio between the average cut ($Cut$) and the optimal cut ($Cut^*$), is depicted for each graph-size in the following scenarios: optimal cut (green), GW (black squares), \qemc noiseless simulation (blue circles), and \qemc real-hardware IBMQ calculations (red stars). The corresponding hyperparameters and exact cuts can be found in Table~\ref{table:hyper_parameters}.}. 
    \label{fig:best_res_4_32}
\end{figure}

\paragraph*{The 4-32 nodes graph instances.} Fig.~\ref{fig:best_res_4_32} provides a summary of the results obtained for the 3-regular graph instances with 4-32 nodes. 
The results are presented for GW (black squares), QEMC noiseless simulations (blue circles), and QEMC real-hardware calculations on IBMQ machines (red stars).  Each method is evaluated based on its averaged cut-ratio, i.e.\ the ratio between the averaged cut and the optimal cut ($Cut/Cut^*$), obtained through exhaustive search. The optimal cut-ratio ($Cut^*/Cut^* = 1$) is indicated by a green, dashed, line. The average for the GW and QEMC classical simulations was computed based on 10 calculations, while for the QEMC real-hardware results, we averaged over 5 runs (except for the 16-node graph instance, which we ran 3 times). 
Table \ref{table:hyper_parameters} provides a summary of these results, along with the corresponding hyperparameters we used.

The results indicate that the average noiseless QEMC simulations achieved the optimal cut for graphs containing up to 14 nodes, and consistently obtained nearly optimal cuts for all graph instances, where 0.9725 was the lowest recorded average cut-ratio, attained for the 30-node graph instance. Additionally, these simulations surpass the average GW results across all tested graph instances. Note that we report the actual average GW results, not the theoretically guaranteed approximation ratio \cite{goemans1995improved_GW},  
which is significantly smaller. 
 Moreover, when focusing on the best execution out of 10 trials, detailed in the ``Max cut" row in Table~\ref{table:hyper_parameters}, it is evident that the top QEMC run consistently reached the optimal exhaustive cut, a feat that was not always met by the GW method.
 Finally, these simulations also exceed the approximation ratio of 0.9326, guaranteed by the enhanced GW-based algorithm, tailored for 3-regular graphs \cite{halperin2004max}.

Fig.~\ref{fig:best_res_4_32} further illustrates that the real-hardware calculations reached the optimal cut for both the 4- and 8-node graph instances, on average. 
For the larger graph instances, with 16 and 32 nodes, the average real-hardware calculations experienced a reduction in the achievable cut, primarily attributed to hardware noise. Here, the constrained number of layers and the actual number of iterations also played a role in limiting the achievable cut. First, larger graph instances often require deeper quantum circuits (see second line in Table.~\ref{table:hyper_parameters} and Sec.~\ref{sec:emp_resource_estimation} for an empirical evaluation of the scaling of the number of layers $L$). However, as mentioned earlier, in an effort to lessen the effect of noise, we used at most two layers in our real-hardware executions. 
Second, due to limited hardware availability, the average  number of actual iterations was 98.6 for the 16-node graphs and 62.0 for the 32-node graphs, as also indicated in Table~\ref{table:hyper_parameters}. 
In addition, a great amount of noise can be attributed to  additional swap-gates introduced to our circuits during  transpilation, in order to make them compatible with the qubit topology of the quantum machines we used. This issue can be mitigated either by using quantum computers with improved qubit connectivity or by opting for slightly less expressive but more hardware-efficient Ans\"atze.

Despite the real-hardware limitations described above, the QEMC real-hardware calculations were still able to achieve cut-ratios of 0.8857 and 0.76 for the 16- and 32-node instances, respectively. Furthermore, Table~\ref{table:hyper_parameters} outlines the maximum cuts reached (according to the best execution), showing that the real-hardware calculations attained top cut-ratios of 0.9 for the 16-node graph instance and 0.81 for the 32-node graph instance. Accomplishing such outcomes for the MaxCut problem using real quantum devices, and especially with superconducting hardware, is no simple task, as evidenced by recent QAOA calculations on Google's quantum processor \cite{harrigan2021quantum}, and IBMQ machines \cite{weidenfeller2022scaling}.

Since the QEMC algorithm was tested on the exact same graph instances as those studied in Ref.~\cite{harrigan2021quantum}, it allows for a direct comparison. To facilitate this, Fig.~\ref{fig:google_scale} in the \sm\ replicates the results of Fig.~\ref{fig:best_res_4_32}, but with a rescaling to align with the evaluation metric applied in \cite{harrigan2021quantum}. Comparing Fig.~\ref{fig:google_scale} with Fig.~4 in Ref.~\cite{harrigan2021quantum}, demonstrates that the cuts achieved by the QEMC real-hardware calculations are substantially better than those reported in \cite{harrigan2021quantum}.

\subsubsection{Ten graph instances per graph-size} \label{sec:small_10_per_size}
In the following, we examine the robustness of the QEMC algorithm with respect to the choice of hyperparameters. To that end, we evaluated a fresh test set of 3-regular graphs with 4-32 nodes (increasing in increments of 2) using classical state-vector simulations. To ensure a representative sample, we generated 10 graph instances per size category resulting in a total of 150 graph instances. 
The evaluation focused on the performance of shallow circuits with $ L = 1$ and $L = 3$ layers, which may not necessarily be optimal. 
For each size category of the graph, we set the optimal step-size per a given value of $L$, as determined previously by our grid-searches on the single graph instances (see Table.~\ref{table:hyper_parameters}), without fine-tuning it any further for the new set of graphs.

\begin{figure}[h!]
    \centering
    \includegraphics[width=0.38\textwidth]{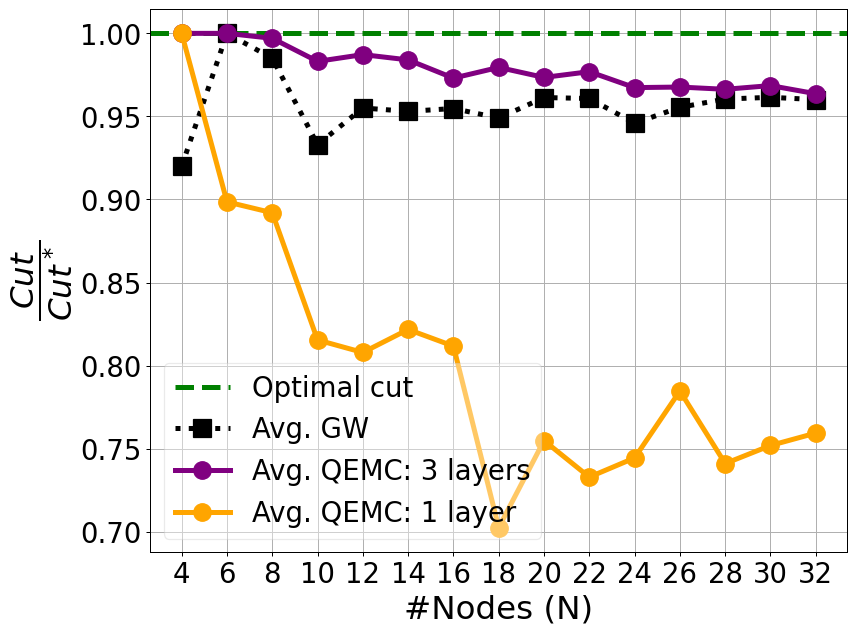}
    \caption{\textbf{Noiseless results for the set of 3-regular graphs with 4 to 32 nodes (2-5 qubits), \textit{ten graph instances per size}.} The average cut-ratio ($\frac{Cut}{Cut^*}$) is depicted for each graph-size in the following cases: optimal cut (green), GW (black squares), \qemc simulations with $L = 3$ circuit-layers (purple circles), and \qemc simulations with $L = 1$ layer (orange circles). Each data-point represents an average of 100 runs, with 10 trials per graph instance.}
    \label{fig:multi_layer_nodes}
\end{figure}

We performed 10 noiseless classical QEMC simulations for each graph instance, using different random initializations of the circuit parameters, as well as 10 GW trials per graph instance. This resulted with a total of 1500 trial runs per method. Fig.~\ref{fig:multi_layer_nodes} presents the average cut-ratios, per graph-size, obtained by: GW (black squares), QEMC 3-layered Ansatz ($L = 3$, purple circles) and QEMC single-layer Ansatz ($L = 1$, orange circles). The results show that the cut-ratios for single-layer Ans\"atze decline rapidly as graph-sizes increase, whereas 3-layered Ans\"atze demonstrate a gradual decrease while consistently achieving average cut-ratios above 0.96 across the entire scale. Comparing these results with the average GW cut-ratios demonstrates that on average, the QEMC algorithm performed better than GW throughout our graphs test set, even when using shallow 3-layered circuits and without explicitly fine-tuning the step-size hyperparameter.

In comparison to QAOA circuits, Ans\"atze with 3 layers are considered fairly shallow. For example, noiseless QAOA simulations
on Erdős–Rényi graphs with 10 nodes required at least 6 layers to outdo GW performance \cite{crooks2018performance_QAOA}. Moreover, each layer in the QAOA circuit demands a number of entangling gates that grows linearly with the number of edges in the graph, leading to a quadratic increase with the number of nodes for general graphs and a linear growth for regular ones. Conversely, in the QEMC algorithm, the number of entangling gates per layer depends on the chosen Ansatz. In the case of the ``strongly-entangling-layers" Ansatz we utilized, it grows logarithmically with the number of nodes, 
resulting in circuit layers that are exponentially shallower than in QAOA. This characteristic of shallow circuits in QEMC enhances its resilience to hardware noise.

\begin{figure*}[t!]
    \centering
    \includegraphics[width=0.8\textwidth]{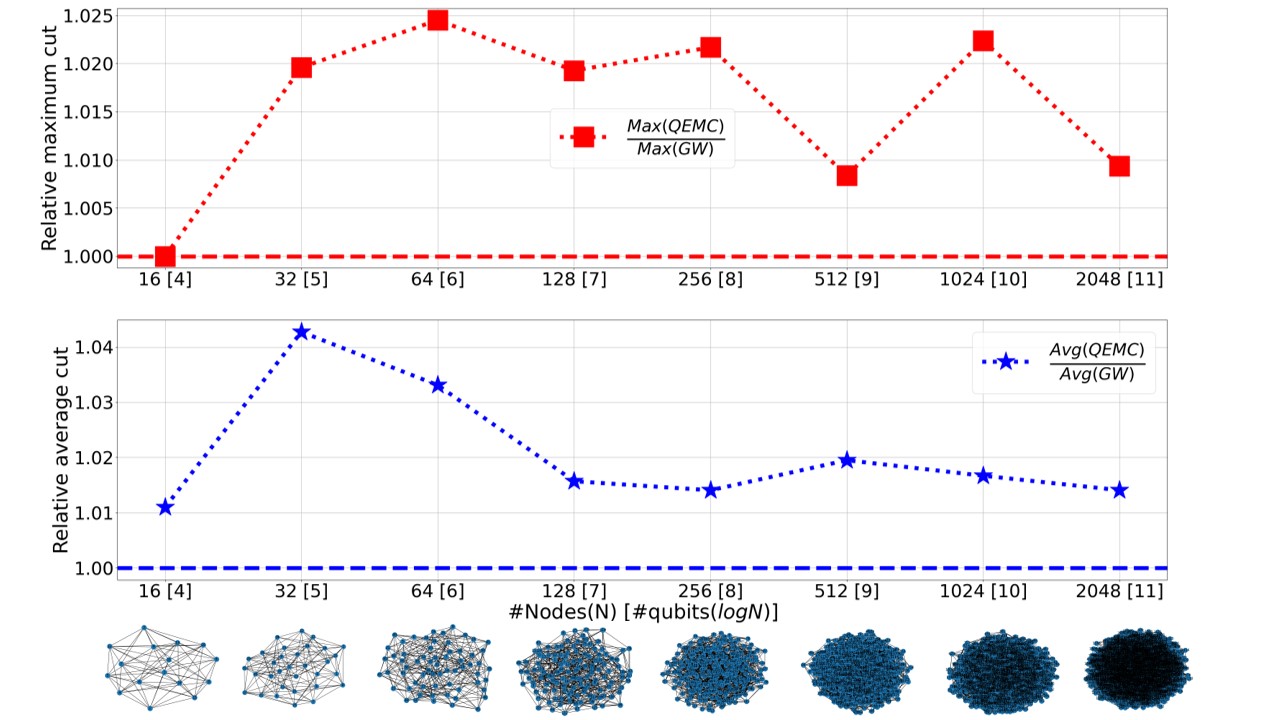}
    \caption{\textbf{Relative performance of the QEMC algorithm (noiseless simulations) compared to the GW algorithm for the  set of 9-regular random graphs, ranging from 16 to 2048 nodes (4-11 qubits in QEMC)}. Each graph-size is represented by a single graph instance, as illustrated at the bottom. \textbf{Top:} The ratio of the best cut found by QEMC to the best cut found by GW, calculated from 10 trials (red squares). \textbf{Bottom:} The ratio of the average QEMC cut to the average GW cut, determined over 10 trials (blue stars).}
    \label{fig:16 to 2048 nodes}
\end{figure*}

\subsection{\label{sec:16-2048 nodes results} Large 9-regular graphs: 16-2048 nodes (4-11 qubits) - noiseless simulations}

The advantages of employing the QEMC encoding scheme become more apparent when examining larger graphs. The logarithmic relationship between the number of required qubits and the size of the graph enables the application of the QEMC method to extensive graphs that would be infeasible to tackle with exhaustive search. In this section, we demonstrate the performance of the QEMC algorithm on 9-regular graphs ranging from 16 to 2048 nodes (increasing in powers of two), which were encoded using 4 to 11 qubits. All QEMC results presented in this section were obtained through classical noiseless state-vector simulations.

\subsubsection{A single graph instance per graph-size}
\label{secsec:16-2048-single}

We randomly generated a single instance of a 9-regular graph for each graph-size (16-2024 nodes), and then determined the optimal hyperparameters for each graph instance using the grid-search approach described in Sec.~\ref{secsec:computational_details}. To that end, we performed $10$ QEMC runs for each combination of graph-size, step-size, and number of layers, where each trial utilized different random circuit parameter initialization, and selected the hyperparameters that resulted with the highest average cut for each graph-size. 
An illustration of the grid-search outcomes for the 256-node graph can be found in 
Fig.~\ref{fig:grid_search_256} in the \sm, and the resulting optimal hyperparameters are listed in Table~\ref{table:hyper_parameters_large}.

We evaluated the performance of QEMC algorithm using optimal hyperparameters by comparing its average and maximum cuts to those of the GW method, obtained over $10$ different runs for each graph instance. Fig.~\ref{fig:16 to 2048 nodes} illustrates the relative performance of QEMC compared to GW. In the top panel, the ratio $\frac{Max(QEMC)}{Max(GW)}$ is depicted (red squares). For the 16-node graph instance, both GW and QEMC achieved the same optimal cut. However, for larger graph instances, the maximum cut obtained by QEMC surpassed the maximum cut obtained by GW by $\approx 1-2.5 \%$. The bottom panel displays the relative average performance of QEMC compared to GW, represented by $\frac{Avg.(QEMC)}{Avg.(GW)}$ (blue stars). It is evident that QEMC outperformed GW on average by $\approx 1-4 \%$ for these graph instances.

\subsubsection{50 random graphs with $N=256$ nodes (8 qubits)}
To further illustrate the strength of the QEMC algorithm and its resilience to hyperparameters selection, we examined its performance on a set of 50 randomly sampled 9-regular graphs, each consisting of $N=256$ nodes (8 qubits), with potentially varying maximum cuts. 
We maintained consistent hyperparameters (step-size and number of layers) across all graph instances, deriving them from the grid-search conducted for the single 256-node graph detailed in Sec.~\ref{secsec:16-2048-single} and illustrated in Fig.~\ref{fig:grid_search_256} in the \sm. To expedite the calculations, we reduced the number of layers to $L=50$ from the optimal value of $L=80$ layers, and adjusted the step-size to $\alpha=0.14$. While not optimal, this configuration resulted in a nearly optimal average cut ($827.2$ instead of $828.5$) in the previously mentioned grid-search. We executed $10$ QEMC runs per graph instance.

\begin{figure}[h!]
    \centering
    \includegraphics[width=0.38\textwidth]{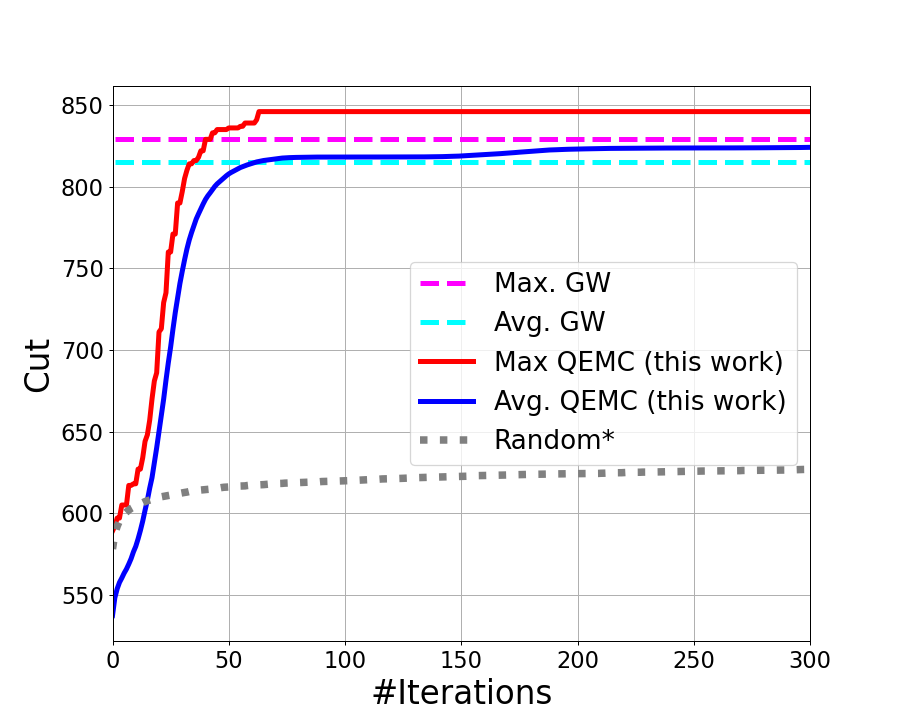}
    \caption{\textbf{QEMC noiseless simulations vs. GW: 50 different 9-regular graph instances with 256 nodes (8 qubits).} \quad \textbf{Max.\ GW (dashed pink line):} Average of the maximum 50 GW cuts, one per graph instance. 
 \textbf{Avg.\ GW (dashed cyan line):} The GW cuts averaged across all 500 trials, 10 trials per graph instance.
 \textbf{Max.\ QEMC (solid red curve):} Average of the maximum 50 best-so-far QEMC cuts, one per graph instance, as a function of QEMC-iterations. 
 \textbf{Avg.\ QEMC (solid blue curve):} The best-so-far QEMC cuts averaged across all 500 trials, 10 trials per graph instance, as a function of QEMC-iterations. We performed 10 trials per graph instance.}
    \label{fig:avg_50_256_graphs}
\end{figure}

Fig.~\ref{fig:avg_50_256_graphs} shows the performance of the \qemc algorithm as a function of QEMC-iterations. 
The Max QEMC curve (solid red line) represents the average over the best cuts achieved for each graph instance (a total number of 50 "best cuts"). It converges to the value of $Max\ QEMC = 846.0$. The average QEMC curve (solid blue line) is simply the average over all the 500 QEMC calculations (50 graph instances, 10 QEMC-runs for each instance). It converges to the value of $Avg.\ QEMC = 824.1$. It is seen that in both cases, the vast majority of the convergence takes place within less than 100 iterations, where the $Max\ QEMC$ curve converges faster.

For comparison, we also executed the GW algorithm on the same set of graphs, running it 10 times per graph instance. The average of the 50 best GW cuts, one per graph instance, yielded a $Max.\ GW$ cut value of 829.1 (dashed pink line), and the $Avg.\ GW$ cut across all trials amounted to 814.9 (dashed cyan line). It is seen that the maximal QEMC curve surpasses the maximal GW cut, crossing it in less than 50 iterations. Moreover, the average QEMC curve outperforms the average GW cut and nearly reaches the the maximal GW cut.

\begin{table*}[t]
\centering
 \begin{tabular}{||l | l|  c c c c c c c c||} 
 \hline
 \multicolumn{2}{||l|}{$\#$ Nodes $(N)$}  & 16 & 32 & 64 & 128 & 256 & 512 & 1024 & 2048  \\ [0.5ex] 
 \hline\hline
  & step-size   & 0.7  & 0.7 &  0.1 & 0.2 & 0.08 &0.02 & 0.04& 0.04 \\
Noiseless & $\#$ layers (to opt.) &  5 & 5 & 50  &40  &80 &120 & 100& 120\\
state-vector & $\#$ layers (to GW) &  2 & 5 &  5 & 20 & 20 & 40 & 100 & 100\\
 simulations & $\#$ iterations (total) & 200  & 200  &  200  & 200 &  200 & 1000  &  1000 & 1000 \\
 & $\#$ iterations (to GW) &  43 & 40  & 30  & 85 &  52 & 560 & 427 & 459 \\
\hline
\hline
Avg. cut & GW  &  45.5 & 98.3 &  199.5 & 407.4  & 817.0 & 1633.8 & 3285.8 & 6580.3 \\
& QEMC simulation& 46    & 102.5  & 206.1    & 413.8    & 828.5 & 1665.6 & 3340.6 & 6673.0 \\
\hline
Max cut & GW &  46 & 102 & 204  & 415 & 830 & 1672 & 3307& 6630\\
& QEMC simulation & 46  & 104 & 209  &  423 & 848 & 1686& 3381&6692 \\
\hline
\hline
 \end{tabular}
 \caption{\textbf{QEMC hyperparameters and attained cuts for the sampled set of 9-regular graphs, ranging from 16 to 2048 nodes, with one graph instance per size}. 
 \quad \textbf{Noiseless state-vector simulations}: The table specifies the optimal step-size and number of layers, as determined via grid-search, the minimal number of layers required to reach the averaged GW cut, the total number of iterations performed, and the iteration count required to reach the averaged GW cut. 
 \textbf{Avg. cut:} The average cuts of GW and QEMC simulations are reported, based on 10 trials per graph instance. 
 \textbf{Max cut:} The maximum cuts from the best trials are reported for GW and QEMC simulations.}
 \label{table:hyper_parameters_large}
\end{table*}

\section{Computational Resources Estimation}
\label{sec:resources}
In what follows, we provide an estimation of the classical and quantum computational resources required for the QEMC algorithm, in terms of the number of nodes ($N = |V|$) and edges ($M = |E|$) in the graph. 

Depending on the choice of the classical optimizer, updating the tunable parameters requires a different number of cost function evaluations. In the worst-case, which we analyze, at each iteration, each parameter update requires its own cost function evaluation(s), e.g.\ in gradient-based optimizers.

Within this framework, the time-to-solution ($TTS$) of the QEMC algorithm is given by:
 \begin{equation}
    \label{eq:TTS}
    TTS =\mathcal{O} \big{[} (T_C+T_Q)\cdot P\cdot I \big{]}
\end{equation}
where $T_C$ and $T_Q$ are the classical and quantum times required for a single cost function evaluation, while $P$ represents the number of tunable parameters in the circuit, and $I$ represents the number of QEMC-iterations.
The actual procedure of the parameters update is considered here to be a constant, subsumed in the  $\mathcal{O}$ notation.

\subsection{QEMC analysis} 
We next analyze the required quantum and classical resources, per cost function evaluation. 

\paragraph{Quantum resources analysis} Memory-wise, the number of qubits ($n$) in the QEMC algorithm is given by $n=\ceil{\log{N}}$, by design. 
Each cost function evaluation requires the calculation of the $N$-dimensional probability distribution, constructed from multiple measurements of the quantum state.
To that end, the quantum processor executes the quantum circuit $S$ times, which is the number of shots needed to achieve a reliable statistical accuracy; Consequently, the quantum circuit execution time per cost function evaluation is  $T_Q = \mathcal{O}(mS)$, where $m$ is the number of gates in the circuit (which is an upper bound on the actual circuit-depth).
The number of gates $m$ in the quantum circuit is Ansatz dependent. In the case of the ``strongly entangling layers" Ansatz we utilized, $m$ is in the order of $\mathcal{O}(L\log{N})$, where $L$ denotes the number of layers in the Ansatz circuit. Therefore, we can conclude that $T_Q = \mathcal{O}(LS \log{N})$, and by neglecting logarithmic terms we arrive at $T_Q = \mathcal{\tilde{O}}(LS)$.

\paragraph{Classical resources analysis} At each cost function evaluation, 
given the $N$-dimensional probability histogram, the classical processor computes the cost function defined in Eq.~\ref{eq:QEMC_cost}, which requires classical time and space that scale linearly with the number of edges, namely $T_C = \mathcal{O}(M)$.
This analysis is summarized in Table.~\ref{table:resources}, under the ``QEMC" row. 

\subsection{QEMC classical simulation} 
Let us now focus on the computational resources required for simulating the QEMC algorithm on classical computers.
We assume a simple ``Schr\"{o}dinger algorithm" \cite{aaronson2016complexity,pednault2017breaking} as a basis for our analysis (further improvements exist, see e.g. \cite{fatima2021faster}). Using this algorithm, simulating a quantum circuit comprising of $n$ qubits and $m$ gates requires $\mathcal{O}(m2^n)$  time and $\mathcal{O}(2^n)$ memory.
As discussed above, we estimate the number of gates $m$ as $\mathcal{O}(L\log{N})$. Consequently, the estimated time complexity for a classical simulation of the QEMC circuit per cost function evaluation is $\mathcal{O}(LN \log{N})$, and by neglecting logarithmic terms we arrive at $\mathcal{\tilde{O}}(LN)$. The space complexity is $\mathcal{O}(N)$.
Taking into account in addition also the time and memory required for calculating the cost function, which scale linearly with the number of edges $M$, we get that simulating the QEMC on a classical computer has  a time complexity of  
$\tilde{\mathcal{O}}(M + LN)$ per a single cost function evaluation, and the memory complexity is $\mathcal{O}(M)$, see Table.~\ref{table:resources}, under the ``\textbf{cs}-QEMC" row.

\renewcommand{\arraystretch}{1.2}
\begin{table}
\begin{tabular}{ |c|c|c|c|c|c| } 
\hline
 \multirow{2}{3em}{   } & \multicolumn{2}{|c|}{Time}  &  \multicolumn{2}{|c|}{Memory} \\
\hline
 & classical ($T_C$) & quantum ($T_Q$) & classical & quantum\\
 \hline
\hline
QEMC & $\mathcal{O}(M)$ & $\tilde{\mathcal{O}}(LS)$ & $\mathcal{O}(M)$ & $\ceil{\log{N}}$ \\ 
\hline
\textit{\textbf{cs}}-QEMC & $\tilde{\mathcal{O}}(M + LN)$ & --- & $\mathcal{O}(M)$ & --- \\
\hline
\end{tabular}
\caption{Computational resources estimation of: 1. \textit{QEMC}; 2. \textit{\textbf{cs}-QEMC} - the classical simulation counterpart of QEMC; The analysis is done per a single cost function evaluation, performed per a single QEMC-iteration, for a fixed size of graph-partition $B$. Note that $N = |V|$ is the number of nodes and $M = |E|$ is the number of edges in the graph, $L$ is the number of layers in the quantum circuit, and $S$ is the number of shots.}
\label{table:resources}
\end{table}

\subsection{Empirical resources estimation} 
\label{sec:emp_resource_estimation}
To gain a clearer understanding of the overall QEMC time-to-solution, we performed an empirical estimation of how the number of layers $(L)$, shots $(S)$, and iterations $(I)$ scale with respect to $N$. Our analysis, conducted through classical state-vector noiseless simulations, aims to provide an empirical upper bound on the computational resources required for the average QEMC cut to achieve the average GW cut, assuming optimal fixation of all other resource parameters. We focused on the same sets of 3-regular and 9-regular graphs introduced in Sec.~\ref{secsec:4-32-single} and ~\ref{secsec:16-2048-single}, respectively, consisting of one graph instance per size, with 10 randomly-initialized executions per graph instance.

\begin{figure*}
    \centering
    \subfloat[\centering Layers]
        {{\includegraphics[width=0.33\textwidth]{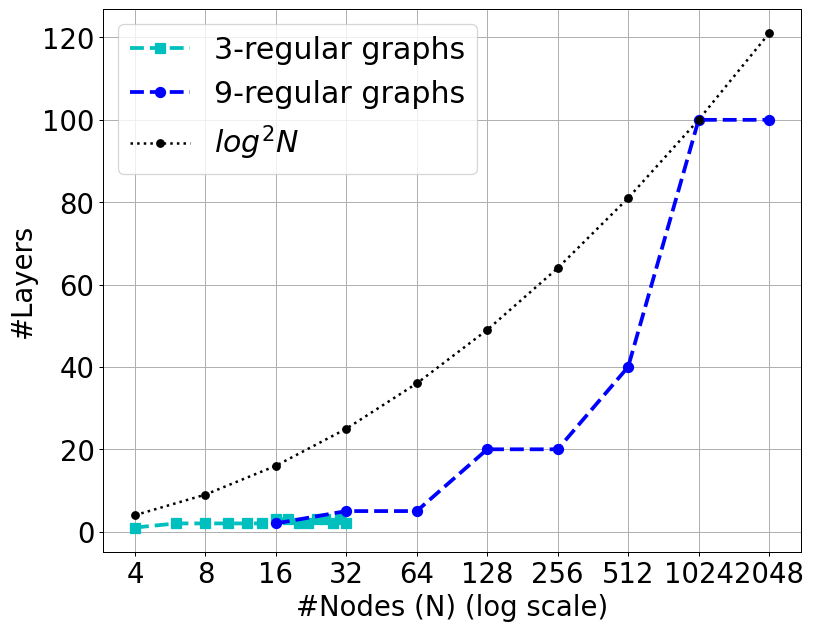} 
        }\label{fig:layers_scaling}}
    \subfloat[\centering Shots]            
        {{\includegraphics[width=0.325\textwidth]{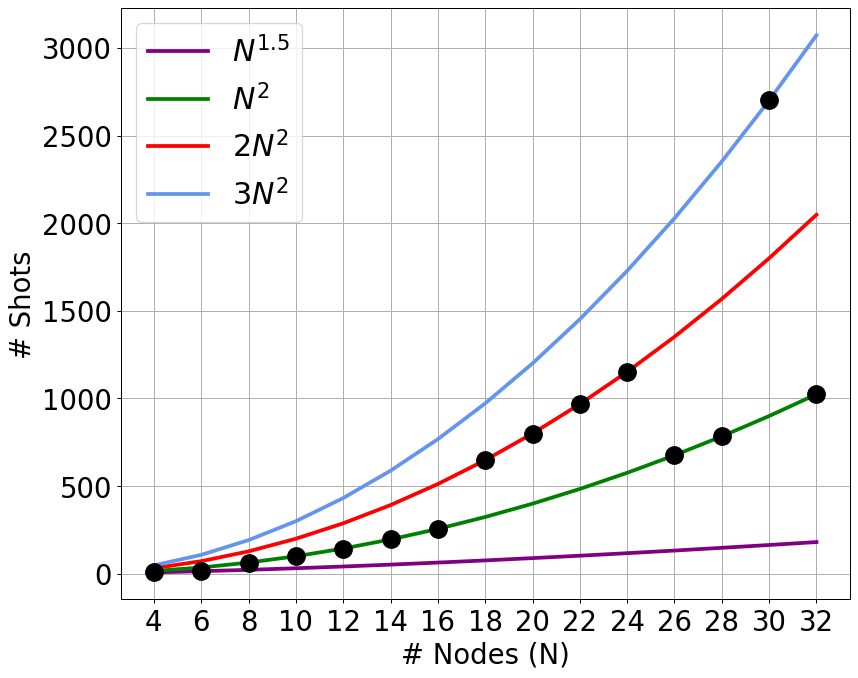} }\label{fig:shots_per_nodes}} 
     \subfloat[\centering Iterations]          
        {{\includegraphics[width=0.33\textwidth]{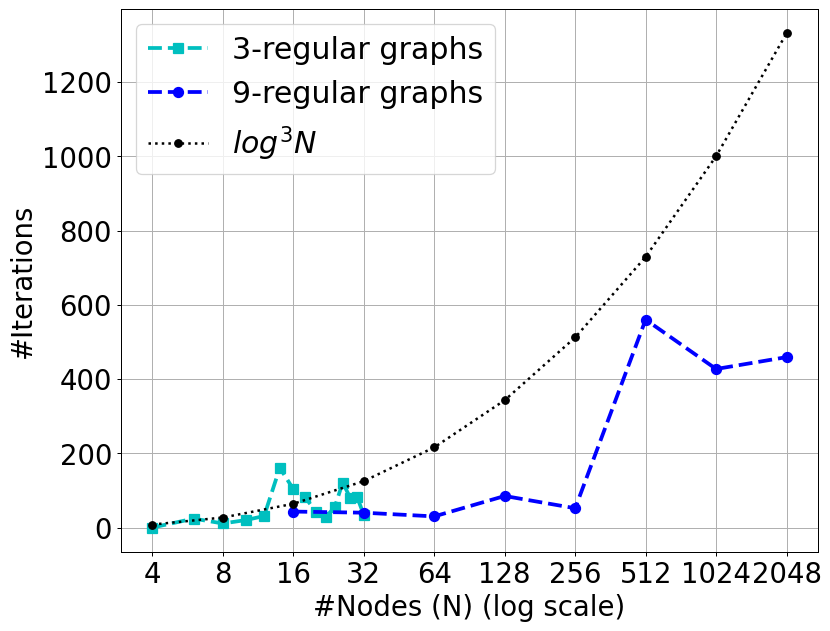} }\label{fig:iterations_scaling}}
     \caption{QEMC resource scaling with the size of the graphs ($N$) required to reach the average GW cut: \textbf{(a)} Scaling of the number of layers ($L$), shown on a logarithmic scale, for the small 4-32 nodes set (in cyan) and the larger 16-2048 nodes set (in blue); \textbf{(b)} Scaling of the number of shots per a single cost function evaluation ($S$), for the small 4-32 nodes set; \textbf{(c)} Scaling of the number of iterations ($I$), displayed on a logarithmic scale, for the small 4-32 nodes set (in cyan) and the larger 16-2048 nodes set (in blue).}
    \label{fig:resources}
\end{figure*}

Fig.~\ref{fig:layers_scaling} shows the minimum number of layers ($L$) required by the QEMC algorithm to reach the GW cut (on average), as a function of $N$, on a logarithmic scale. 
The minimum number of layers was determined through the hyperparameters grid-searches described in Sec.~\ref{sec:results}, using infinite number of shots and the same total number of iterations reported for the classical simulations.
To get an estimation of the scaling we further plot a $\log^2{N}$ curve that upper bounds the results, suggesting a poly-logarithmic scaling of $L$ with respect to $N$. 

Next, Fig.~\ref{fig:shots_per_nodes} illustrates the scaling of the number of shots per iteration ($S$) with respect to $N$. We conducted QEMC simulations using varying shot quantities per cost function evaluation, namely $N,N^{3/2},N^2,2N^2$, and $3N^2$. Our focus was on the set of 3-regular graphs with smaller sizes ranging from 4 to 32 nodes, as finite shot-number simulations are more resource-intensive. Throughout the simulations we utilized the optimal hyperparameters (number of layers, step-size) detailed in Table~\ref{table:hyper_parameters} and allowed for the same total number of iterations reported for the classical simulations ($I=300$). We marked the minimum shot quantity at which the average GW cut was reached. The results depicted in Fig.~\ref{fig:shots_per_nodes} suggest an approximated quadratic scaling of $S$ with $N$.

Finally, Fig.~\ref{fig:iterations_scaling} shows the number of QEMC-iterations ($I$) required in order to reach the GW cut (on average), as a function of $N$, on a logarithmic scale. 
We simulated each graph instance with the optimal hyperparameters reported in Tables \ref{table:hyper_parameters}-\ref{table:hyper_parameters_large} for the classical simulations, using infinite number of shots. Here, too, we further plotted a $\log^3{N}$ curve that upper bounds the results, suggesting a poly-logarithmic scaling of the number of the iterations $I$ with $N$.

Overall, our empirical estimation indicates that $L$, and therefore also $P$ scale as $ \mathcal{O}(\polylog{N})$, while $S\sim \mathcal{O}(N^2)$, and $I\sim \mathcal{O}(\polylog{N})$. Therefore, it can be deduced, empirically-based, that the QEMC time-to-solution ($TTS$, Eq. \ref{eq:TTS}) is governed by the number of required shots ($S$), and scales quadratically with the number of nodes, i.e.\ $\tilde{\mathcal{O}}(N^2)$ when  neglecting logarithmic terms. 

Our analysis assumes that the size of the graph-partition, represented by the value of $B$, can be estimated to a good accuracy in advance. If this isn't feasible, the scheme would need to be executed $\lfloor\frac{N}{2}\rfloor$ times, leading to a total time-to-solution which scales cubically with the number of nodes $N$. 
This positions the QEMC algorithm as a polynomial-time heuristic solver for the MaxCut problem.
In comparison, note that the mere specification of the graph scales linearly with the number of edges $M$, which is in general quadratic with the number of nodes $N$. In addition, the time complexity of the GW algorithm is $\tilde{\mathcal{O}}(N^{3.5})$ in the most general cases, $\tilde{\mathcal{O}}(NM)$, when all weights are non-negative, and $\tilde{\mathcal{O}}(M) = \tilde{\mathcal{O}}(N)$ for regular graphs, see \cite{haribara2016coherent} and references therein.

Finally, according to our empirical findings, it seems that no speedup is provided by the QEMC algorithm ($\mathcal{\tilde{O}}(M + LS)$), compared to its classical counterpart \textbf{cs}-QEMC ($\mathcal{\tilde{O}}(M + LN)$). This entails that the QEMC algorithm, in its current formulation, does not provide a straightforward quantum advantage. Instead, it offers a quantum-inspired, efficient heuristic for solving the MaxCut problem. 

\section{Conclusions and Discussions}
\label{sec:conclusions_discussions}
To conclude, we presented a novel variational quantum algorithm designed to heuristically address the \mcp for $N$-node graphs, using  $\ceil{\log{N}}$ qubits. 
The QEMC algorithm utilizes a unique encoding scheme that maps the logical state of a node (either ``blue" or ``white") to a range of probability values for measuring the corresponding computational state, by means of probability-threshold encoding. 
The performance of our algorithm was numerically evaluated using both real-hardware IBMQ calculations and classical simulations, focusing on a set of unweighted, regular-graph instances. 

For graph instances with up to $32$ nodes, our real-hardware computations attained 
cuts that were considerably higher compared to previous studies using QAOA. 
We ascribe QEMC's enhanced performance on quantum hardware over QAOA to two key factors: first, its employment of significantly smaller circuits, which reduces noise susceptibility;
and second, its threshold probability encoding scheme. This encoding approach allows the algorithm to associate each graph partition with a volume of quantum states.
This, in turn, grants the variational process a broader optimal state space to explore, potentially enhancing the system's noise resilience. 

In addition, our noiseless QEMC simulations, extended to graph instances of up to $2048$ nodes, consistently demonstrated an edge over the GW algorithm. Following this, an empirical resources evaluation suggested that the QEMC algorithm is computationally efficient, requiring time and memory resources that scale approximately quadratically with the graph-size $(N)$. While these results are encouraging, it is important to note that they do not challenge the theoretical bounds described by \cite{khot2007optimal_uniqe_game_conjecture_GW_optimal}, as they are based on empirical tests over a fixed set of graph instances and do not guarantee a theoretically-proven minimal approximation ratio exceeding that of the GW algorithm.

Based on an empirical analysis of the required computational resources for both QEMC and its classically-simulated counterpart (\textbf{cs}-QEMC), we have shown that, as expected, the QEMC algorithm does not provide a quantum advantage. 
Demonstrating such an advantage requires an approach that not only surpasses traditional classical schemes but also outperforms the algorithm’s own classical simulation, a merit that is more challenging for qubit-efficient VQAs.

Nevertheless, the QEMC algorithm may serve as a competitive quantum-inspired classical algorithm. This novel paradigm holds potential for numerous applications in classical computing. Moreover, to the best of our knowledge, this is the first quantum-inspired algorithm based on a variational quantum approach. As such, it stands as a challenging benchmark for comparing state-of-the-art variational quantum algorithms, such as QAOA and its derivatives, in terms of computational resources (circuit depth, number of shots, and number of iterations) and overall performance. With respect to circuit depth, we have already made empirical observations that indicate that the QEMC algorithm necessitates circuits that are exponentially shallower than those used in QAOA, in order to surpass the GW algorithm's performance.

To date, most physically-inspired, classical and quantum techniques towards combinatorial problems, including simulated annealing, coherent Ising machines, quantum annealing and the QAOA, formalize the problem naturally in terms of an Ising Hamiltonian \cite{mohseni2022ising}. Our formulation is different and phrases the problem as a continuous optimization problem in a form that may be applicable also for other types of optimization problems. This highlights the potential significance of exploring new quantum information encoding schemes.

Several open questions remain: (1) Are there any settings under which the QEMC, or variations of, could also offer quantum advantage? For example, in cases where the reconstruction of the entire probability distribution is not required? (2) What insights can be drawn about the potential quantum advantage of QAOA, by comparing it to our QEMC algorithm?
(3) While GW is the best-known approximation algorithm for the MaxCut problem, guaranteeing the highest cut in the worst-case scenario, it may not always be the best choice in practice \cite{burer2002rank, haribara2016coherent}. How does the QEMC algorithm perform in comparison to state-of-the-art heuristics like simulated annealing \cite{1983_Kirkpatrick_Science}, rank-two relaxation heuristic \cite{burer2002rank}, and the coherent Ising machine \cite{honjo2021100}? (4) We have tested the QEMC algorithm only on unweighted regular graphs, how does it perform on other families of graphs? These questions are left for future research.

\section{Acknowledgments}
We acknowledge the use of IBM Quantum services for this work. The views expressed are those of the authors, and do not reflect the official policy or position of IBM or the IBM Quantum team.

\bibliography{apssamp}

\providecommand{\noopsort}[1]{}\providecommand{\singleletter}[1]{#1}%
\begin{thebibliography}{52}%
\makeatletter
\providecommand \@ifxundefined [1]{%
 \@ifx{#1\undefined}
}%
\providecommand \@ifnum [1]{%
 \ifnum #1\expandafter \@firstoftwo
 \else \expandafter \@secondoftwo
 \fi
}%
\providecommand \@ifx [1]{%
 \ifx #1\expandafter \@firstoftwo
 \else \expandafter \@secondoftwo
 \fi
}%
\providecommand \natexlab [1]{#1}%
\providecommand \enquote  [1]{``#1''}%
\providecommand \bibnamefont  [1]{#1}%
\providecommand \bibfnamefont [1]{#1}%
\providecommand \citenamefont [1]{#1}%
\providecommand \href@noop [0]{\@secondoftwo}%
\providecommand \href [0]{\begingroup \@sanitize@url \@href}%
\providecommand \@href[1]{\@@startlink{#1}\@@href}%
\providecommand \@@href[1]{\endgroup#1\@@endlink}%
\providecommand \@sanitize@url [0]{\catcode `\\12\catcode `\$12\catcode `\&12\catcode `\#12\catcode `\^12\catcode `\_12\catcode `\%12\relax}%
\providecommand \@@startlink[1]{}%
\providecommand \@@endlink[0]{}%
\providecommand \url  [0]{\begingroup\@sanitize@url \@url }%
\providecommand \@url [1]{\endgroup\@href {#1}{\urlprefix }}%
\providecommand \urlprefix  [0]{URL }%
\providecommand \Eprint [0]{\href }%
\providecommand \doibase [0]{https://doi.org/}%
\providecommand \selectlanguage [0]{\@gobble}%
\providecommand \bibinfo  [0]{\@secondoftwo}%
\providecommand \bibfield  [0]{\@secondoftwo}%
\providecommand \translation [1]{[#1]}%
\providecommand \BibitemOpen [0]{}%
\providecommand \bibitemStop [0]{}%
\providecommand \bibitemNoStop [0]{.\EOS\space}%
\providecommand \EOS [0]{\spacefactor3000\relax}%
\providecommand \BibitemShut  [1]{\csname bibitem#1\endcsname}%
\let\auto@bib@innerbib\@empty
\bibitem [{\citenamefont {Nielsen}\ and\ \citenamefont {Chuang}(2011)}]{2011_Nielsen_Xhuang_QI}%
  \BibitemOpen
  \bibfield  {author} {\bibinfo {author} {\bibfnamefont {M.~A.}\ \bibnamefont {Nielsen}}\ and\ \bibinfo {author} {\bibfnamefont {I.~L.}\ \bibnamefont {Chuang}},\ }\href {https://doi.org/10.1017/CBO9780511976667} {\emph {\bibinfo {title} {Quantum Computation and Quantum Information}}},\ \bibinfo {edition} {10th}\ ed.\ (\bibinfo  {publisher} {Cambridge University Press},\ \bibinfo {address} {New York, NY, USA},\ \bibinfo {year} {2011})\BibitemShut {NoStop}%
\bibitem [{\citenamefont {Preskill}(2018)}]{2018_Quantum_Preskill_NISQ}%
  \BibitemOpen
  \bibfield  {author} {\bibinfo {author} {\bibfnamefont {J.}~\bibnamefont {Preskill}},\ }\bibfield  {title} {\bibinfo {title} {Quantum computing in the {NISQ} era and beyond},\ }\href {https://doi.org/https://doi.org/10.22331/q-2018-08-06-79} {\bibfield  {journal} {\bibinfo  {journal} {Quantum}\ }\textbf {\bibinfo {volume} {2}},\ \bibinfo {pages} {79} (\bibinfo {year} {2018})}\BibitemShut {NoStop}%
\bibitem [{\citenamefont {Peruzzo}\ \emph {et~al.}(2014)\citenamefont {Peruzzo}, \citenamefont {McClean}, \citenamefont {Shadbolt}, \citenamefont {Yung}, \citenamefont {Zhou}, \citenamefont {Love}, \citenamefont {Aspuru-Guzik},\ and\ \citenamefont {O’brien}}]{peruzzo2014variational}%
  \BibitemOpen
  \bibfield  {author} {\bibinfo {author} {\bibfnamefont {A.}~\bibnamefont {Peruzzo}}, \bibinfo {author} {\bibfnamefont {J.}~\bibnamefont {McClean}}, \bibinfo {author} {\bibfnamefont {P.}~\bibnamefont {Shadbolt}}, \bibinfo {author} {\bibfnamefont {M.-H.}\ \bibnamefont {Yung}}, \bibinfo {author} {\bibfnamefont {X.-Q.}\ \bibnamefont {Zhou}}, \bibinfo {author} {\bibfnamefont {P.~J.}\ \bibnamefont {Love}}, \bibinfo {author} {\bibfnamefont {A.}~\bibnamefont {Aspuru-Guzik}},\ and\ \bibinfo {author} {\bibfnamefont {J.~L.}\ \bibnamefont {O’brien}},\ }\bibfield  {title} {\bibinfo {title} {A variational eigenvalue solver on a photonic quantum processor},\ }\href {https://doi.org/https://doi.org/10.1038/ncomms5213} {\bibfield  {journal} {\bibinfo  {journal} {Nature communications}\ }\textbf {\bibinfo {volume} {5}},\ \bibinfo {pages} {4213} (\bibinfo {year} {2014})}\BibitemShut {NoStop}%
\bibitem [{\citenamefont {Tilly}\ \emph {et~al.}(2022)\citenamefont {Tilly}, \citenamefont {Chen}, \citenamefont {Cao}, \citenamefont {Picozzi}, \citenamefont {Setia}, \citenamefont {Li}, \citenamefont {Grant}, \citenamefont {Wossnig}, \citenamefont {Rungger}, \citenamefont {Booth} \emph {et~al.}}]{tilly2022variational}%
  \BibitemOpen
  \bibfield  {author} {\bibinfo {author} {\bibfnamefont {J.}~\bibnamefont {Tilly}}, \bibinfo {author} {\bibfnamefont {H.}~\bibnamefont {Chen}}, \bibinfo {author} {\bibfnamefont {S.}~\bibnamefont {Cao}}, \bibinfo {author} {\bibfnamefont {D.}~\bibnamefont {Picozzi}}, \bibinfo {author} {\bibfnamefont {K.}~\bibnamefont {Setia}}, \bibinfo {author} {\bibfnamefont {Y.}~\bibnamefont {Li}}, \bibinfo {author} {\bibfnamefont {E.}~\bibnamefont {Grant}}, \bibinfo {author} {\bibfnamefont {L.}~\bibnamefont {Wossnig}}, \bibinfo {author} {\bibfnamefont {I.}~\bibnamefont {Rungger}}, \bibinfo {author} {\bibfnamefont {G.~H.}\ \bibnamefont {Booth}}, \emph {et~al.},\ }\bibfield  {title} {\bibinfo {title} {The variational quantum eigensolver: a review of methods and best practices},\ }\href {https://doi.org/https://doi.org/10.1016/j.physrep.2022.08.003} {\bibfield  {journal} {\bibinfo  {journal} {Physics Reports}\ }\textbf {\bibinfo {volume} {986}},\ \bibinfo {pages} {1} (\bibinfo {year} {2022})}\BibitemShut {NoStop}%
\bibitem [{\citenamefont {Mangini}(2023)}]{mangini2023variational}%
  \BibitemOpen
  \bibfield  {author} {\bibinfo {author} {\bibfnamefont {S.}~\bibnamefont {Mangini}},\ }\href@noop {} {\bibinfo {title} {Variational quantum algorithms for machine learning: theory and applications}} (\bibinfo {year} {2023}),\ \Eprint {https://arxiv.org/abs/2306.09984} {arXiv:2306.09984 [quant-ph]} \BibitemShut {NoStop}%
\bibitem [{\citenamefont {Bravo-Prieto}\ \emph {et~al.}(2020)\citenamefont {Bravo-Prieto}, \citenamefont {LaRose}, \citenamefont {Cerezo}, \citenamefont {Subasi}, \citenamefont {Cincio},\ and\ \citenamefont {Coles}}]{bravo2020variational}%
  \BibitemOpen
  \bibfield  {author} {\bibinfo {author} {\bibfnamefont {C.}~\bibnamefont {Bravo-Prieto}}, \bibinfo {author} {\bibfnamefont {R.}~\bibnamefont {LaRose}}, \bibinfo {author} {\bibfnamefont {M.}~\bibnamefont {Cerezo}}, \bibinfo {author} {\bibfnamefont {Y.}~\bibnamefont {Subasi}}, \bibinfo {author} {\bibfnamefont {L.}~\bibnamefont {Cincio}},\ and\ \bibinfo {author} {\bibfnamefont {P.~J.}\ \bibnamefont {Coles}},\ }\href@noop {} {\bibinfo {title} {Variational quantum linear solver}} (\bibinfo {year} {2020}),\ \Eprint {https://arxiv.org/abs/1909.05820} {arXiv:1909.05820 [quant-ph]} \BibitemShut {NoStop}%
\bibitem [{\citenamefont {Farhi}\ \emph {et~al.}(2014)\citenamefont {Farhi}, \citenamefont {Goldstone},\ and\ \citenamefont {Gutmann}}]{farhi2014quantum_org_qaoa}%
  \BibitemOpen
  \bibfield  {author} {\bibinfo {author} {\bibfnamefont {E.}~\bibnamefont {Farhi}}, \bibinfo {author} {\bibfnamefont {J.}~\bibnamefont {Goldstone}},\ and\ \bibinfo {author} {\bibfnamefont {S.}~\bibnamefont {Gutmann}},\ }\href@noop {} {\bibinfo {title} {A quantum approximate optimization algorithm}} (\bibinfo {year} {2014}),\ \Eprint {https://arxiv.org/abs/1411.4028} {arXiv:1411.4028 [quant-ph]} \BibitemShut {NoStop}%
\bibitem [{\citenamefont {Crooks}(2018)}]{crooks2018performance_QAOA}%
  \BibitemOpen
  \bibfield  {author} {\bibinfo {author} {\bibfnamefont {G.~E.}\ \bibnamefont {Crooks}},\ }\href@noop {} {\bibinfo {title} {Performance of the quantum approximate optimization algorithm on the maximum cut problem}} (\bibinfo {year} {2018}),\ \Eprint {https://arxiv.org/abs/1811.08419} {arXiv:1811.08419 [quant-ph]} \BibitemShut {NoStop}%
\bibitem [{\citenamefont {Zhou}\ \emph {et~al.}(2020)\citenamefont {Zhou}, \citenamefont {Wang}, \citenamefont {Choi}, \citenamefont {Pichler},\ and\ \citenamefont {Lukin}}]{2020_PRX_Lukin_QAOA_performance}%
  \BibitemOpen
  \bibfield  {author} {\bibinfo {author} {\bibfnamefont {L.}~\bibnamefont {Zhou}}, \bibinfo {author} {\bibfnamefont {S.-T.}\ \bibnamefont {Wang}}, \bibinfo {author} {\bibfnamefont {S.}~\bibnamefont {Choi}}, \bibinfo {author} {\bibfnamefont {H.}~\bibnamefont {Pichler}},\ and\ \bibinfo {author} {\bibfnamefont {M.~D.}\ \bibnamefont {Lukin}},\ }\bibfield  {title} {\bibinfo {title} {Quantum approximate optimization algorithm: performance, mechanism, and implementation on near-term devices},\ }\href {https://doi.org/10.1103/PhysRevX.10.021067} {\bibfield  {journal} {\bibinfo  {journal} {Phys. Rev. X}\ }\textbf {\bibinfo {volume} {10}},\ \bibinfo {pages} {021067} (\bibinfo {year} {2020})}\BibitemShut {NoStop}%
\bibitem [{\citenamefont {Wurtz}\ and\ \citenamefont {Lykov}(2021)}]{wurtz2021fixed}%
  \BibitemOpen
  \bibfield  {author} {\bibinfo {author} {\bibfnamefont {J.}~\bibnamefont {Wurtz}}\ and\ \bibinfo {author} {\bibfnamefont {D.}~\bibnamefont {Lykov}},\ }\bibfield  {title} {\bibinfo {title} {Fixed-angle conjectures for the quantum approximate optimization algorithm on regular {MaxCut} graphs},\ }\href {https://doi.org/10.1103/PhysRevA.104.052419} {\bibfield  {journal} {\bibinfo  {journal} {Phys. Rev. A}\ }\textbf {\bibinfo {volume} {104}},\ \bibinfo {pages} {052419} (\bibinfo {year} {2021})}\BibitemShut {NoStop}%
\bibitem [{\citenamefont {Harrigan}\ \emph {et~al.}(2021)\citenamefont {Harrigan}, \citenamefont {Sung}, \citenamefont {Neeley}, \citenamefont {Satzinger}, \citenamefont {Arute}, \citenamefont {Arya}, \citenamefont {Atalaya}, \citenamefont {Bardin}, \citenamefont {Barends}, \citenamefont {Boixo} \emph {et~al.}}]{harrigan2021quantum}%
  \BibitemOpen
  \bibfield  {author} {\bibinfo {author} {\bibfnamefont {M.~P.}\ \bibnamefont {Harrigan}}, \bibinfo {author} {\bibfnamefont {K.~J.}\ \bibnamefont {Sung}}, \bibinfo {author} {\bibfnamefont {M.}~\bibnamefont {Neeley}}, \bibinfo {author} {\bibfnamefont {K.~J.}\ \bibnamefont {Satzinger}}, \bibinfo {author} {\bibfnamefont {F.}~\bibnamefont {Arute}}, \bibinfo {author} {\bibfnamefont {K.}~\bibnamefont {Arya}}, \bibinfo {author} {\bibfnamefont {J.}~\bibnamefont {Atalaya}}, \bibinfo {author} {\bibfnamefont {J.~C.}\ \bibnamefont {Bardin}}, \bibinfo {author} {\bibfnamefont {R.}~\bibnamefont {Barends}}, \bibinfo {author} {\bibfnamefont {S.}~\bibnamefont {Boixo}}, \emph {et~al.},\ }\bibfield  {title} {\bibinfo {title} {Quantum approximate optimization of non-planar graph problems on a planar superconducting processor},\ }\href {https://doi.org/10.1038/s41567-020-01105-y} {\bibfield  {journal} {\bibinfo  {journal} {Nature Physics}\ }\textbf {\bibinfo {volume} {17}},\ \bibinfo {pages} {332} (\bibinfo {year}
  {2021})}\BibitemShut {NoStop}%
\bibitem [{\citenamefont {Weidenfeller}\ \emph {et~al.}(2022)\citenamefont {Weidenfeller}, \citenamefont {Valor}, \citenamefont {Gacon}, \citenamefont {Tornow}, \citenamefont {Bello}, \citenamefont {Woerner},\ and\ \citenamefont {Egger}}]{weidenfeller2022scaling}%
  \BibitemOpen
  \bibfield  {author} {\bibinfo {author} {\bibfnamefont {J.}~\bibnamefont {Weidenfeller}}, \bibinfo {author} {\bibfnamefont {L.~C.}\ \bibnamefont {Valor}}, \bibinfo {author} {\bibfnamefont {J.}~\bibnamefont {Gacon}}, \bibinfo {author} {\bibfnamefont {C.}~\bibnamefont {Tornow}}, \bibinfo {author} {\bibfnamefont {L.}~\bibnamefont {Bello}}, \bibinfo {author} {\bibfnamefont {S.}~\bibnamefont {Woerner}},\ and\ \bibinfo {author} {\bibfnamefont {D.~J.}\ \bibnamefont {Egger}},\ }\bibfield  {title} {\bibinfo {title} {Scaling of the quantum approximate optimization algorithm on superconducting qubit based hardware},\ }\href {https://doi.org/https://doi.org/10.22331/q-2022-12-07-870} {\bibfield  {journal} {\bibinfo  {journal} {Quantum}\ }\textbf {\bibinfo {volume} {6}},\ \bibinfo {pages} {870} (\bibinfo {year} {2022})}\BibitemShut {NoStop}%
\bibitem [{\citenamefont {Guerrero}(2020)}]{Guerrero2020SolvingCO}%
  \BibitemOpen
  \bibfield  {author} {\bibinfo {author} {\bibfnamefont {N.~J.}\ \bibnamefont {Guerrero}},\ }\emph {\bibinfo {title} {{Solving combinatorial optimization problems using the quantum approximation optimization algorithm}}},\ \href {https://scholar.afit.edu/etd/3263} {\bibinfo {type} {Theses and dissertations}},\ \bibinfo  {school} {Air Force Institute of Technology} (\bibinfo {year} {2020})\BibitemShut {NoStop}%
\bibitem [{\citenamefont {Guerreschi}\ and\ \citenamefont {Matsuura}(2019)}]{guerreschi2019qaoa}%
  \BibitemOpen
  \bibfield  {author} {\bibinfo {author} {\bibfnamefont {G.~G.}\ \bibnamefont {Guerreschi}}\ and\ \bibinfo {author} {\bibfnamefont {A.~Y.}\ \bibnamefont {Matsuura}},\ }\bibfield  {title} {\bibinfo {title} {{QAOA for Max-Cut requires hundreds of qubits for quantum speed-up}},\ }\href {https://doi.org/10.1038/s41598-019-43176-9} {\bibfield  {journal} {\bibinfo  {journal} {Scientific Reports}\ }\textbf {\bibinfo {volume} {9}},\ \bibinfo {pages} {6903} (\bibinfo {year} {2019})}\BibitemShut {NoStop}%
\bibitem [{\citenamefont {Dunjko}\ \emph {et~al.}(2018)\citenamefont {Dunjko}, \citenamefont {Ge},\ and\ \citenamefont {Cirac}}]{dunjko2018computational}%
  \BibitemOpen
  \bibfield  {author} {\bibinfo {author} {\bibfnamefont {V.}~\bibnamefont {Dunjko}}, \bibinfo {author} {\bibfnamefont {Y.}~\bibnamefont {Ge}},\ and\ \bibinfo {author} {\bibfnamefont {J.~I.}\ \bibnamefont {Cirac}},\ }\bibfield  {title} {\bibinfo {title} {Computational speedups using small quantum devices},\ }\href {https://doi.org/10.1103/PhysRevLett.121.250501} {\bibfield  {journal} {\bibinfo  {journal} {Phys. Rev. Lett.}\ }\textbf {\bibinfo {volume} {121}},\ \bibinfo {pages} {250501} (\bibinfo {year} {2018})}\BibitemShut {NoStop}%
\bibitem [{\citenamefont {Ge}\ and\ \citenamefont {Dunjko}(2020)}]{ge2020hybrid}%
  \BibitemOpen
  \bibfield  {author} {\bibinfo {author} {\bibfnamefont {Y.}~\bibnamefont {Ge}}\ and\ \bibinfo {author} {\bibfnamefont {V.}~\bibnamefont {Dunjko}},\ }\bibfield  {title} {\bibinfo {title} {{A hybrid algorithm framework for small quantum computers with application to finding Hamiltonian cycles}},\ }\href {https://doi.org/10.1063/1.5119235} {\bibfield  {journal} {\bibinfo  {journal} {Journal of Mathematical Physics}\ }\textbf {\bibinfo {volume} {61}},\ \bibinfo {pages} {012201} (\bibinfo {year} {2020})}\BibitemShut {NoStop}%
\bibitem [{\citenamefont {Guerreschi}(2021)}]{guerreschi2021solving}%
  \BibitemOpen
  \bibfield  {author} {\bibinfo {author} {\bibfnamefont {G.~G.}\ \bibnamefont {Guerreschi}},\ }\href@noop {} {\bibinfo {title} {{Solving quadratic unconstrained binary optimization with divide-and-conquer and quantum algorithms}}} (\bibinfo {year} {2021}),\ \Eprint {https://arxiv.org/abs/2101.07813} {arXiv:2101.07813 [quant-ph]} \BibitemShut {NoStop}%
\bibitem [{\citenamefont {Li}\ \emph {et~al.}(2023)\citenamefont {Li}, \citenamefont {Alam},\ and\ \citenamefont {Ghosh}}]{li2022large}%
  \BibitemOpen
  \bibfield  {author} {\bibinfo {author} {\bibfnamefont {J.}~\bibnamefont {Li}}, \bibinfo {author} {\bibfnamefont {M.}~\bibnamefont {Alam}},\ and\ \bibinfo {author} {\bibfnamefont {S.}~\bibnamefont {Ghosh}},\ }\bibfield  {title} {\bibinfo {title} {{Large-scale quantum approximate optimization via divide-and-conquer}},\ }\href {https://doi.org/10.1109/TCAD.2022.3212196} {\bibfield  {journal} {\bibinfo  {journal} {IEEE Transactions on Computer-Aided Design of Integrated Circuits and Systems}\ }\textbf {\bibinfo {volume} {42}},\ \bibinfo {pages} {1852} (\bibinfo {year} {2023})}\BibitemShut {NoStop}%
\bibitem [{\citenamefont {Zhou}\ \emph {et~al.}(2023)\citenamefont {Zhou}, \citenamefont {Du}, \citenamefont {Tian},\ and\ \citenamefont {Tao}}]{zhou2022qaoa_in_qaoa}%
  \BibitemOpen
  \bibfield  {author} {\bibinfo {author} {\bibfnamefont {Z.}~\bibnamefont {Zhou}}, \bibinfo {author} {\bibfnamefont {Y.}~\bibnamefont {Du}}, \bibinfo {author} {\bibfnamefont {X.}~\bibnamefont {Tian}},\ and\ \bibinfo {author} {\bibfnamefont {D.}~\bibnamefont {Tao}},\ }\bibfield  {title} {\bibinfo {title} {{QAOA-in-QAOA: Solving large-scale MaxCut problems on small quantum machines}},\ }\href {https://doi.org/10.1103/PhysRevApplied.19.024027} {\bibfield  {journal} {\bibinfo  {journal} {Phys. Rev. Appl.}\ }\textbf {\bibinfo {volume} {19}},\ \bibinfo {pages} {024027} (\bibinfo {year} {2023})}\BibitemShut {NoStop}%
\bibitem [{\citenamefont {Glos}\ \emph {et~al.}(2022)\citenamefont {Glos}, \citenamefont {Krawiec},\ and\ \citenamefont {Zimborás}}]{glos2022space}%
  \BibitemOpen
  \bibfield  {author} {\bibinfo {author} {\bibfnamefont {A.}~\bibnamefont {Glos}}, \bibinfo {author} {\bibfnamefont {A.}~\bibnamefont {Krawiec}},\ and\ \bibinfo {author} {\bibfnamefont {Z.}~\bibnamefont {Zimborás}},\ }\bibfield  {title} {\bibinfo {title} {Space-efficient binary optimization for variational quantum computing},\ }\href {https://doi.org/10.1038/s41534-022-00546-y} {\bibfield  {journal} {\bibinfo  {journal} {npj Quantum Information}\ }\textbf {\bibinfo {volume} {8}},\ \bibinfo {pages} {39} (\bibinfo {year} {2022})}\BibitemShut {NoStop}%
\bibitem [{\citenamefont {Tan}\ \emph {et~al.}(2021)\citenamefont {Tan}, \citenamefont {Lemonde}, \citenamefont {Thanasilp}, \citenamefont {Tangpanitanon},\ and\ \citenamefont {Angelakis}}]{tan2021qubit}%
  \BibitemOpen
  \bibfield  {author} {\bibinfo {author} {\bibfnamefont {B.}~\bibnamefont {Tan}}, \bibinfo {author} {\bibfnamefont {M.-A.}\ \bibnamefont {Lemonde}}, \bibinfo {author} {\bibfnamefont {S.}~\bibnamefont {Thanasilp}}, \bibinfo {author} {\bibfnamefont {J.}~\bibnamefont {Tangpanitanon}},\ and\ \bibinfo {author} {\bibfnamefont {D.~G.}\ \bibnamefont {Angelakis}},\ }\bibfield  {title} {\bibinfo {title} {{Qubit-efficient encoding schemes for binary optimisation problems}},\ }\href {https://doi.org/10.22331/q-2021-05-04-454} {\bibfield  {journal} {\bibinfo  {journal} {{Quantum}}\ }\textbf {\bibinfo {volume} {5}},\ \bibinfo {pages} {454} (\bibinfo {year} {2021})}\BibitemShut {NoStop}%
\bibitem [{\citenamefont {Tabi}\ \emph {et~al.}(2020)\citenamefont {Tabi}, \citenamefont {El-Safty}, \citenamefont {Kallus}, \citenamefont {Hága}, \citenamefont {Kozsik}, \citenamefont {Glos},\ and\ \citenamefont {Zimborás}}]{tabi2020quantum}%
  \BibitemOpen
  \bibfield  {author} {\bibinfo {author} {\bibfnamefont {Z.}~\bibnamefont {Tabi}}, \bibinfo {author} {\bibfnamefont {K.~H.}\ \bibnamefont {El-Safty}}, \bibinfo {author} {\bibfnamefont {Z.}~\bibnamefont {Kallus}}, \bibinfo {author} {\bibfnamefont {P.}~\bibnamefont {Hága}}, \bibinfo {author} {\bibfnamefont {T.}~\bibnamefont {Kozsik}}, \bibinfo {author} {\bibfnamefont {A.}~\bibnamefont {Glos}},\ and\ \bibinfo {author} {\bibfnamefont {Z.}~\bibnamefont {Zimborás}},\ }\bibfield  {title} {\bibinfo {title} {Quantum optimization for the graph coloring problem with space-efficient embedding},\ }in\ \href {https://doi.org/10.1109/QCE49297.2020.00018} {\emph {\bibinfo {booktitle} {2020 IEEE International Conference on Quantum Computing and Engineering (QCE)}}}\ (\bibinfo {year} {2020})\ pp.\ \bibinfo {pages} {56--62}\BibitemShut {NoStop}%
\bibitem [{\citenamefont {Fuchs}\ \emph {et~al.}(2021)\citenamefont {Fuchs}, \citenamefont {Kolden}, \citenamefont {Aase},\ and\ \citenamefont {Sartor}}]{fuchs2021efficient}%
  \BibitemOpen
  \bibfield  {author} {\bibinfo {author} {\bibfnamefont {F.~G.}\ \bibnamefont {Fuchs}}, \bibinfo {author} {\bibfnamefont {H.~{\O}.}\ \bibnamefont {Kolden}}, \bibinfo {author} {\bibfnamefont {N.~H.}\ \bibnamefont {Aase}},\ and\ \bibinfo {author} {\bibfnamefont {G.}~\bibnamefont {Sartor}},\ }\bibfield  {title} {\bibinfo {title} {Efficient encoding of the weighted max k-cut on a quantum computer using qaoa},\ }\href {https://doi.org/10.1007/s42979-020-00437-z} {\bibfield  {journal} {\bibinfo  {journal} {SN Computer Science}\ }\textbf {\bibinfo {volume} {2}},\ \bibinfo {pages} {89} (\bibinfo {year} {2021})}\BibitemShut {NoStop}%
\bibitem [{\citenamefont {Patti}\ \emph {et~al.}(2022)\citenamefont {Patti}, \citenamefont {Kossaifi}, \citenamefont {Anandkumar},\ and\ \citenamefont {Yelin}}]{patti2022variational}%
  \BibitemOpen
  \bibfield  {author} {\bibinfo {author} {\bibfnamefont {T.~L.}\ \bibnamefont {Patti}}, \bibinfo {author} {\bibfnamefont {J.}~\bibnamefont {Kossaifi}}, \bibinfo {author} {\bibfnamefont {A.}~\bibnamefont {Anandkumar}},\ and\ \bibinfo {author} {\bibfnamefont {S.~F.}\ \bibnamefont {Yelin}},\ }\bibfield  {title} {\bibinfo {title} {Variational quantum optimization with multibasis encodings},\ }\href {https://doi.org/10.1103/PhysRevResearch.4.033142} {\bibfield  {journal} {\bibinfo  {journal} {Phys. Rev. Res.}\ }\textbf {\bibinfo {volume} {4}},\ \bibinfo {pages} {033142} (\bibinfo {year} {2022})}\BibitemShut {NoStop}%
\bibitem [{\citenamefont {Ran\ifmmode \check{c}\else \v{c}\fi{}i\ifmmode~\acute{c}\else \'{c}\fi{}}(2023)}]{ranvcic2023noisy}%
  \BibitemOpen
  \bibfield  {author} {\bibinfo {author} {\bibfnamefont {M.~J.}\ \bibnamefont {Ran\ifmmode \check{c}\else \v{c}\fi{}i\ifmmode~\acute{c}\else \'{c}\fi{}}},\ }\bibfield  {title} {\bibinfo {title} {{Noisy intermediate-scale quantum computing algorithm for solving an $n$-vertex MaxCut problem with log($n$) qubits}},\ }\href {https://doi.org/10.1103/PhysRevResearch.5.L012021} {\bibfield  {journal} {\bibinfo  {journal} {Phys. Rev. Res.}\ }\textbf {\bibinfo {volume} {5}},\ \bibinfo {pages} {L012021} (\bibinfo {year} {2023})}\BibitemShut {NoStop}%
\bibitem [{\citenamefont {Chatterjee}\ \emph {et~al.}(2023)\citenamefont {Chatterjee}, \citenamefont {Bourreau},\ and\ \citenamefont {Rančić}}]{chatterjee2023solving}%
  \BibitemOpen
  \bibfield  {author} {\bibinfo {author} {\bibfnamefont {Y.}~\bibnamefont {Chatterjee}}, \bibinfo {author} {\bibfnamefont {E.}~\bibnamefont {Bourreau}},\ and\ \bibinfo {author} {\bibfnamefont {M.~J.}\ \bibnamefont {Rančić}},\ }\href@noop {} {\bibinfo {title} {{Solving various NP-Hard problems using exponentially fewer qubits on a quantum computer}}} (\bibinfo {year} {2023}),\ \Eprint {https://arxiv.org/abs/2301.06978} {arXiv:2301.06978 [quant-ph]} \BibitemShut {NoStop}%
\bibitem [{\citenamefont {Aaronson}(2015)}]{aaronson2015read}%
  \BibitemOpen
  \bibfield  {author} {\bibinfo {author} {\bibfnamefont {S.}~\bibnamefont {Aaronson}},\ }\bibfield  {title} {\bibinfo {title} {Read the fine print},\ }\href@noop {} {\bibfield  {journal} {\bibinfo  {journal} {Nature Physics}\ }\textbf {\bibinfo {volume} {11}},\ \bibinfo {pages} {291} (\bibinfo {year} {2015})}\BibitemShut {NoStop}%
\bibitem [{\citenamefont {Karp}(1972)}]{Karp1972}%
  \BibitemOpen
  \bibfield  {author} {\bibinfo {author} {\bibfnamefont {R.~M.}\ \bibnamefont {Karp}},\ }\bibinfo {title} {{Reducibility among combinatorial problems}},\ in\ \href {https://doi.org/10.1007/978-1-4684-2001-2_9} {\emph {\bibinfo {booktitle} {Complexity of Computer Computations: Proceedings of a symposium on the Complexity of Computer Computations, held March 20--22, 1972, at the IBM Thomas J. Watson Research Center, Yorktown Heights, New York, and sponsored by the Office of Naval Research, Mathematics Program, IBM World Trade Corporation, and the IBM Research Mathematical Sciences Department}}},\ \bibinfo {editor} {edited by\ \bibinfo {editor} {\bibfnamefont {R.~E.}\ \bibnamefont {Miller}}, \bibinfo {editor} {\bibfnamefont {J.~W.}\ \bibnamefont {Thatcher}},\ and\ \bibinfo {editor} {\bibfnamefont {J.~D.}\ \bibnamefont {Bohlinger}}}\ (\bibinfo  {publisher} {Springer US},\ \bibinfo {address} {Boston, MA},\ \bibinfo {year} {1972})\ pp.\ \bibinfo {pages} {85--103}\BibitemShut {NoStop}%
\bibitem [{\citenamefont {Barahona}\ \emph {et~al.}(1988)\citenamefont {Barahona}, \citenamefont {Grötschel}, \citenamefont {Jünger},\ and\ \citenamefont {Reinelt}}]{AnApplicationofCombinatorialOptimizationtoStatisticalPhysicsandCircuitLayoutDesign}%
  \BibitemOpen
  \bibfield  {author} {\bibinfo {author} {\bibfnamefont {F.}~\bibnamefont {Barahona}}, \bibinfo {author} {\bibfnamefont {M.}~\bibnamefont {Grötschel}}, \bibinfo {author} {\bibfnamefont {M.}~\bibnamefont {Jünger}},\ and\ \bibinfo {author} {\bibfnamefont {G.}~\bibnamefont {Reinelt}},\ }\bibfield  {title} {\bibinfo {title} {{An application of combinatorial optimization to statistical physics and circuit layout design}},\ }\href {https://doi.org/10.1287/opre.36.3.493} {\bibfield  {journal} {\bibinfo  {journal} {Operations Research}\ }\textbf {\bibinfo {volume} {36}},\ \bibinfo {pages} {493} (\bibinfo {year} {1988})}\BibitemShut {NoStop}%
\bibitem [{\citenamefont {Goemans}\ and\ \citenamefont {Williamson}(1995)}]{goemans1995improved_GW}%
  \BibitemOpen
  \bibfield  {author} {\bibinfo {author} {\bibfnamefont {M.~X.}\ \bibnamefont {Goemans}}\ and\ \bibinfo {author} {\bibfnamefont {D.~P.}\ \bibnamefont {Williamson}},\ }\bibfield  {title} {\bibinfo {title} {Improved approximation algorithms for maximum cut and satisfiability problems using semidefinite programming},\ }\href {https://doi.org/10.1145/227683.227684} {\bibfield  {journal} {\bibinfo  {journal} {Journal of the ACM (JACM)}\ }\textbf {\bibinfo {volume} {42}},\ \bibinfo {pages} {1115} (\bibinfo {year} {1995})}\BibitemShut {NoStop}%
\bibitem [{\citenamefont {Kandala}\ \emph {et~al.}(2017)\citenamefont {Kandala}, \citenamefont {Mezzacapo}, \citenamefont {Temme}, \citenamefont {Takita}, \citenamefont {Brink}, \citenamefont {Chow},\ and\ \citenamefont {Gambetta}}]{kandala2017hardware}%
  \BibitemOpen
  \bibfield  {author} {\bibinfo {author} {\bibfnamefont {A.}~\bibnamefont {Kandala}}, \bibinfo {author} {\bibfnamefont {A.}~\bibnamefont {Mezzacapo}}, \bibinfo {author} {\bibfnamefont {K.}~\bibnamefont {Temme}}, \bibinfo {author} {\bibfnamefont {M.}~\bibnamefont {Takita}}, \bibinfo {author} {\bibfnamefont {M.}~\bibnamefont {Brink}}, \bibinfo {author} {\bibfnamefont {J.~M.}\ \bibnamefont {Chow}},\ and\ \bibinfo {author} {\bibfnamefont {J.~M.}\ \bibnamefont {Gambetta}},\ }\bibfield  {title} {\bibinfo {title} {Hardware-efficient variational quantum eigensolver for small molecules and quantum magnets},\ }\href {https://doi.org/10.1038/nature23879} {\bibfield  {journal} {\bibinfo  {journal} {nature}\ }\textbf {\bibinfo {volume} {549}},\ \bibinfo {pages} {242} (\bibinfo {year} {2017})}\BibitemShut {NoStop}%
\bibitem [{\citenamefont {Meitei}\ \emph {et~al.}(2021)\citenamefont {Meitei}, \citenamefont {Gard}, \citenamefont {Barron}, \citenamefont {Pappas}, \citenamefont {Economou}, \citenamefont {Barnes},\ and\ \citenamefont {Mayhall}}]{meitei2021gate}%
  \BibitemOpen
  \bibfield  {author} {\bibinfo {author} {\bibfnamefont {O.~R.}\ \bibnamefont {Meitei}}, \bibinfo {author} {\bibfnamefont {B.~T.}\ \bibnamefont {Gard}}, \bibinfo {author} {\bibfnamefont {G.~S.}\ \bibnamefont {Barron}}, \bibinfo {author} {\bibfnamefont {D.~P.}\ \bibnamefont {Pappas}}, \bibinfo {author} {\bibfnamefont {S.~E.}\ \bibnamefont {Economou}}, \bibinfo {author} {\bibfnamefont {E.}~\bibnamefont {Barnes}},\ and\ \bibinfo {author} {\bibfnamefont {N.~J.}\ \bibnamefont {Mayhall}},\ }\bibfield  {title} {\bibinfo {title} {Gate-free state preparation for fast variational quantum eigensolver simulations},\ }\href {https://doi.org/https://doi.org/10.1038/s41534-021-00493-0} {\bibfield  {journal} {\bibinfo  {journal} {npj Quantum Information}\ }\textbf {\bibinfo {volume} {7}},\ \bibinfo {pages} {155} (\bibinfo {year} {2021})}\BibitemShut {NoStop}%
\bibitem [{\citenamefont {Liang}\ \emph {et~al.}(2023)\citenamefont {Liang}, \citenamefont {Cheng}, \citenamefont {Ren}, \citenamefont {Wang}, \citenamefont {Hua}, \citenamefont {Song}, \citenamefont {Ding}, \citenamefont {Chong}, \citenamefont {Han}, \citenamefont {Shi},\ and\ \citenamefont {Qian}}]{liang2023napa}%
  \BibitemOpen
  \bibfield  {author} {\bibinfo {author} {\bibfnamefont {Z.}~\bibnamefont {Liang}}, \bibinfo {author} {\bibfnamefont {J.}~\bibnamefont {Cheng}}, \bibinfo {author} {\bibfnamefont {H.}~\bibnamefont {Ren}}, \bibinfo {author} {\bibfnamefont {H.}~\bibnamefont {Wang}}, \bibinfo {author} {\bibfnamefont {F.}~\bibnamefont {Hua}}, \bibinfo {author} {\bibfnamefont {Z.}~\bibnamefont {Song}}, \bibinfo {author} {\bibfnamefont {Y.}~\bibnamefont {Ding}}, \bibinfo {author} {\bibfnamefont {F.}~\bibnamefont {Chong}}, \bibinfo {author} {\bibfnamefont {S.}~\bibnamefont {Han}}, \bibinfo {author} {\bibfnamefont {Y.}~\bibnamefont {Shi}},\ and\ \bibinfo {author} {\bibfnamefont {X.}~\bibnamefont {Qian}},\ }\href@noop {} {\bibinfo {title} {Napa: Intermediate-level variational native-pulse ansatz for variational quantum algorithms}} (\bibinfo {year} {2023}),\ \Eprint {https://arxiv.org/abs/2208.01215} {arXiv:2208.01215 [quant-ph]} \BibitemShut {NoStop}%
\bibitem [{\citenamefont {Meirom}\ and\ \citenamefont {Frankel}(2022)}]{meirom2022pansatz}%
  \BibitemOpen
  \bibfield  {author} {\bibinfo {author} {\bibfnamefont {D.}~\bibnamefont {Meirom}}\ and\ \bibinfo {author} {\bibfnamefont {S.~H.}\ \bibnamefont {Frankel}},\ }\href@noop {} {\bibinfo {title} {{PANSATZ: Pulse-based ansatz for variational quantum algorithms}}} (\bibinfo {year} {2022}),\ \Eprint {https://arxiv.org/abs/2212.12911} {arXiv:2212.12911 [quant-ph]} \BibitemShut {NoStop}%
\bibitem [{\citenamefont {De~Keijzer}\ \emph {et~al.}(2023)\citenamefont {De~Keijzer}, \citenamefont {Tse},\ and\ \citenamefont {Kokkelmans}}]{de2023pulse}%
  \BibitemOpen
  \bibfield  {author} {\bibinfo {author} {\bibfnamefont {R.}~\bibnamefont {De~Keijzer}}, \bibinfo {author} {\bibfnamefont {O.}~\bibnamefont {Tse}},\ and\ \bibinfo {author} {\bibfnamefont {S.}~\bibnamefont {Kokkelmans}},\ }\bibfield  {title} {\bibinfo {title} {Pulse based variational quantum optimal control for hybrid quantum computing},\ }\href {https://doi.org/https://doi.org/10.22331/q-2023-01-26-908} {\bibfield  {journal} {\bibinfo  {journal} {Quantum}\ }\textbf {\bibinfo {volume} {7}},\ \bibinfo {pages} {908} (\bibinfo {year} {2023})}\BibitemShut {NoStop}%
\bibitem [{\citenamefont {Gorb}()}]{rvg77_maxcut}%
  \BibitemOpen
  \bibfield  {author} {\bibinfo {author} {\bibfnamefont {R.}~\bibnamefont {Gorb}},\ }\href@noop {} {\bibinfo {title} {An implementation of the \text{G}oemans-\text{W}illiamson algorithm}},\ \bibinfo {howpublished} {\url{https://github.com/rvg77/max-cut/blob/master/src/models/maxcut.py}}\BibitemShut {NoStop}%
\bibitem [{\citenamefont {Schuld}\ \emph {et~al.}(2020)\citenamefont {Schuld}, \citenamefont {Bocharov}, \citenamefont {Svore},\ and\ \citenamefont {Wiebe}}]{schuld2020circuit}%
  \BibitemOpen
  \bibfield  {author} {\bibinfo {author} {\bibfnamefont {M.}~\bibnamefont {Schuld}}, \bibinfo {author} {\bibfnamefont {A.}~\bibnamefont {Bocharov}}, \bibinfo {author} {\bibfnamefont {K.~M.}\ \bibnamefont {Svore}},\ and\ \bibinfo {author} {\bibfnamefont {N.}~\bibnamefont {Wiebe}},\ }\bibfield  {title} {\bibinfo {title} {Circuit-centric quantum classifiers},\ }\href {https://doi.org/10.1103/physreva.101.032308} {\bibfield  {journal} {\bibinfo  {journal} {Physical Review A}\ }\textbf {\bibinfo {volume} {101}},\ \bibinfo {pages} {032308} (\bibinfo {year} {2020})}\BibitemShut {NoStop}%
\bibitem [{\citenamefont {Bergholm}\ \emph {et~al.}(2020)\citenamefont {Bergholm}, \citenamefont {Izaac}, \citenamefont {Schuld}, \citenamefont {Gogolin}, \citenamefont {Ahmed}, \citenamefont {Ajith}, \citenamefont {Alam}, \citenamefont {Alonso-Linaje}, \citenamefont {AkashNarayanan}, \citenamefont {Asadi} \emph {et~al.}}]{Pennylane}%
  \BibitemOpen
  \bibfield  {author} {\bibinfo {author} {\bibfnamefont {V.}~\bibnamefont {Bergholm}}, \bibinfo {author} {\bibfnamefont {J.}~\bibnamefont {Izaac}}, \bibinfo {author} {\bibfnamefont {M.}~\bibnamefont {Schuld}}, \bibinfo {author} {\bibfnamefont {C.}~\bibnamefont {Gogolin}}, \bibinfo {author} {\bibfnamefont {S.}~\bibnamefont {Ahmed}}, \bibinfo {author} {\bibfnamefont {V.}~\bibnamefont {Ajith}}, \bibinfo {author} {\bibfnamefont {M.~S.}\ \bibnamefont {Alam}}, \bibinfo {author} {\bibfnamefont {G.}~\bibnamefont {Alonso-Linaje}}, \bibinfo {author} {\bibfnamefont {B.}~\bibnamefont {AkashNarayanan}}, \bibinfo {author} {\bibfnamefont {A.}~\bibnamefont {Asadi}}, \emph {et~al.},\ }\href@noop {} {\bibinfo {title} {Pennylane: Automatic differentiation of hybrid quantum-classical computations}} (\bibinfo {year} {2020}),\ \Eprint {https://arxiv.org/abs/1811.04968} {arXiv:1811.04968 [quant-ph]} \BibitemShut {NoStop}%
\bibitem [{\citenamefont {Sack}\ and\ \citenamefont {Serbyn}(2021)}]{sack2021quantum}%
  \BibitemOpen
  \bibfield  {author} {\bibinfo {author} {\bibfnamefont {S.~H.}\ \bibnamefont {Sack}}\ and\ \bibinfo {author} {\bibfnamefont {M.}~\bibnamefont {Serbyn}},\ }\bibfield  {title} {\bibinfo {title} {Quantum annealing initialization of the quantum approximate optimization algorithm},\ }\href {https://doi.org/10.22331/q-2021-07-01-491} {\bibfield  {journal} {\bibinfo  {journal} {Quantum}\ }\textbf {\bibinfo {volume} {5}},\ \bibinfo {pages} {491} (\bibinfo {year} {2021})}\BibitemShut {NoStop}%
\bibitem [{\citenamefont {Amosy}\ \emph {et~al.}(2022)\citenamefont {Amosy}, \citenamefont {Danzig}, \citenamefont {Porat}, \citenamefont {Chechik},\ and\ \citenamefont {Makmal}}]{amosy2022iterative}%
  \BibitemOpen
  \bibfield  {author} {\bibinfo {author} {\bibfnamefont {O.}~\bibnamefont {Amosy}}, \bibinfo {author} {\bibfnamefont {T.}~\bibnamefont {Danzig}}, \bibinfo {author} {\bibfnamefont {E.}~\bibnamefont {Porat}}, \bibinfo {author} {\bibfnamefont {G.}~\bibnamefont {Chechik}},\ and\ \bibinfo {author} {\bibfnamefont {A.}~\bibnamefont {Makmal}},\ }\href@noop {} {\bibinfo {title} {Iterative-free quantum approximate optimization algorithm using neural networks}} (\bibinfo {year} {2022}),\ \Eprint {https://arxiv.org/abs/2208.09888} {arXiv:2208.09888 [quant-ph]} \BibitemShut {NoStop}%
\bibitem [{\citenamefont {Kingma}\ and\ \citenamefont {Ba}(2014)}]{kingma2014adam}%
  \BibitemOpen
  \bibfield  {author} {\bibinfo {author} {\bibfnamefont {D.~P.}\ \bibnamefont {Kingma}}\ and\ \bibinfo {author} {\bibfnamefont {J.}~\bibnamefont {Ba}},\ }\href@noop {} {\bibinfo {title} {{Adam: A method for stochastic optimization}}} (\bibinfo {year} {2014}),\ \Eprint {https://arxiv.org/abs/1412.6980} {arXiv:1412.6980 [cs.LG]} \BibitemShut {NoStop}%
\bibitem [{\citenamefont {Hagberg}\ \emph {et~al.}(2008)\citenamefont {Hagberg}, \citenamefont {Schult},\ and\ \citenamefont {Swart}}]{networkx}%
  \BibitemOpen
  \bibfield  {author} {\bibinfo {author} {\bibfnamefont {A.~A.}\ \bibnamefont {Hagberg}}, \bibinfo {author} {\bibfnamefont {D.~A.}\ \bibnamefont {Schult}},\ and\ \bibinfo {author} {\bibfnamefont {P.~J.}\ \bibnamefont {Swart}},\ }\href@noop {} {\bibinfo {title} {{NetworkX}}},\ \bibinfo {howpublished} {\url{https://networkx.github.io}} (\bibinfo {year} {2008}),\ \bibinfo {note} {{Accessed: 2023-04-17}}\BibitemShut {NoStop}%
\bibitem [{\citenamefont {Halperin}\ \emph {et~al.}(2004)\citenamefont {Halperin}, \citenamefont {Livnat},\ and\ \citenamefont {Zwick}}]{halperin2004max}%
  \BibitemOpen
  \bibfield  {author} {\bibinfo {author} {\bibfnamefont {E.}~\bibnamefont {Halperin}}, \bibinfo {author} {\bibfnamefont {D.}~\bibnamefont {Livnat}},\ and\ \bibinfo {author} {\bibfnamefont {U.}~\bibnamefont {Zwick}},\ }\bibfield  {title} {\bibinfo {title} {{MAX CUT in cubic graphs}},\ }\href {https://doi.org/https://doi.org/10.1016/j.jalgor.2004.06.001} {\bibfield  {journal} {\bibinfo  {journal} {Journal of Algorithms}\ }\textbf {\bibinfo {volume} {53}},\ \bibinfo {pages} {169} (\bibinfo {year} {2004})}\BibitemShut {NoStop}%
\bibitem [{\citenamefont {Aaronson}\ and\ \citenamefont {Chen}(2016)}]{aaronson2016complexity}%
  \BibitemOpen
  \bibfield  {author} {\bibinfo {author} {\bibfnamefont {S.}~\bibnamefont {Aaronson}}\ and\ \bibinfo {author} {\bibfnamefont {L.}~\bibnamefont {Chen}},\ }\href@noop {} {\bibinfo {title} {{Complexity-theoretic foundations of quantum supremacy qxperiments}}} (\bibinfo {year} {2016}),\ \Eprint {https://arxiv.org/abs/1612.05903} {arXiv:1612.05903 [quant-ph]} \BibitemShut {NoStop}%
\bibitem [{\citenamefont {Pednault}\ \emph {et~al.}(2017)\citenamefont {Pednault}, \citenamefont {Gunnels}, \citenamefont {Nannicini}, \citenamefont {Horesh}, \citenamefont {Magerlein}, \citenamefont {Solomonik},\ and\ \citenamefont {Wisnieff}}]{pednault2017breaking}%
  \BibitemOpen
  \bibfield  {author} {\bibinfo {author} {\bibfnamefont {E.}~\bibnamefont {Pednault}}, \bibinfo {author} {\bibfnamefont {J.~A.}\ \bibnamefont {Gunnels}}, \bibinfo {author} {\bibfnamefont {G.}~\bibnamefont {Nannicini}}, \bibinfo {author} {\bibfnamefont {L.}~\bibnamefont {Horesh}}, \bibinfo {author} {\bibfnamefont {T.}~\bibnamefont {Magerlein}}, \bibinfo {author} {\bibfnamefont {E.}~\bibnamefont {Solomonik}},\ and\ \bibinfo {author} {\bibfnamefont {R.}~\bibnamefont {Wisnieff}},\ }\href@noop {} {\bibinfo {title} {Breaking the 49-qubit barrier in the simulation of quantum circuits}} (\bibinfo {year} {2017}),\ \Eprint {https://arxiv.org/abs/1710.05867v1} {arXiv:1710.05867v1 [quant-ph]} \BibitemShut {NoStop}%
\bibitem [{\citenamefont {Fatima}\ and\ \citenamefont {Markov}(2021)}]{fatima2021faster}%
  \BibitemOpen
  \bibfield  {author} {\bibinfo {author} {\bibfnamefont {A.}~\bibnamefont {Fatima}}\ and\ \bibinfo {author} {\bibfnamefont {I.~L.}\ \bibnamefont {Markov}},\ }\bibfield  {title} {\bibinfo {title} {Faster schr{\"o}dinger-style simulation of quantum circuits},\ }in\ \href {https://doi.org/https://doi.org/10.1109/HPCA51647.2021.00026} {\emph {\bibinfo {booktitle} {2021 IEEE International Symposium on High-Performance Computer Architecture (HPCA)}}}\ (\bibinfo {organization} {IEEE},\ \bibinfo {year} {2021})\ pp.\ \bibinfo {pages} {194--207}\BibitemShut {NoStop}%
\bibitem [{\citenamefont {Haribara}\ \emph {et~al.}(2016)\citenamefont {Haribara}, \citenamefont {Utsunomiya},\ and\ \citenamefont {Yamamoto}}]{haribara2016coherent}%
  \BibitemOpen
  \bibfield  {author} {\bibinfo {author} {\bibfnamefont {Y.}~\bibnamefont {Haribara}}, \bibinfo {author} {\bibfnamefont {S.}~\bibnamefont {Utsunomiya}},\ and\ \bibinfo {author} {\bibfnamefont {Y.}~\bibnamefont {Yamamoto}},\ }\bibfield  {title} {\bibinfo {title} {{A coherent Ising machine for MAX-CUT problems: performance evaluation against semidefinite programming and simulated annealing}},\ }\href {https://doi.org/10.1007/978-4-431-55756-2_12} {\bibfield  {journal} {\bibinfo  {journal} {Principles and Methods of Quantum Information Technologies}\ ,\ \bibinfo {pages} {251}} (\bibinfo {year} {2016})}\BibitemShut {NoStop}%
\bibitem [{\citenamefont {Khot}\ \emph {et~al.}(2007)\citenamefont {Khot}, \citenamefont {Kindler}, \citenamefont {Mossel},\ and\ \citenamefont {O’Donnell}}]{khot2007optimal_uniqe_game_conjecture_GW_optimal}%
  \BibitemOpen
  \bibfield  {author} {\bibinfo {author} {\bibfnamefont {S.}~\bibnamefont {Khot}}, \bibinfo {author} {\bibfnamefont {G.}~\bibnamefont {Kindler}}, \bibinfo {author} {\bibfnamefont {E.}~\bibnamefont {Mossel}},\ and\ \bibinfo {author} {\bibfnamefont {R.}~\bibnamefont {O’Donnell}},\ }\bibfield  {title} {\bibinfo {title} {{Optimal inapproximability results for MAX‐CUT and Other 2‐Variable CSPs?}},\ }\href {https://doi.org/10.1137/S0097539705447372} {\bibfield  {journal} {\bibinfo  {journal} {SIAM Journal on Computing}\ }\textbf {\bibinfo {volume} {37}},\ \bibinfo {pages} {319} (\bibinfo {year} {2007})}\BibitemShut {NoStop}%
\bibitem [{\citenamefont {Mohseni}\ \emph {et~al.}(2022)\citenamefont {Mohseni}, \citenamefont {McMahon},\ and\ \citenamefont {Byrnes}}]{mohseni2022ising}%
  \BibitemOpen
  \bibfield  {author} {\bibinfo {author} {\bibfnamefont {N.}~\bibnamefont {Mohseni}}, \bibinfo {author} {\bibfnamefont {P.~L.}\ \bibnamefont {McMahon}},\ and\ \bibinfo {author} {\bibfnamefont {T.}~\bibnamefont {Byrnes}},\ }\bibfield  {title} {\bibinfo {title} {Ising machines as hardware solvers of combinatorial optimization problems},\ }\href {https://doi.org/10.1038/s42254-022-00440-8} {\bibfield  {journal} {\bibinfo  {journal} {Nature Reviews Physics}\ }\textbf {\bibinfo {volume} {4}},\ \bibinfo {pages} {363} (\bibinfo {year} {2022})}\BibitemShut {NoStop}%
\bibitem [{\citenamefont {Burer}\ \emph {et~al.}(2002)\citenamefont {Burer}, \citenamefont {Monteiro},\ and\ \citenamefont {Zhang}}]{burer2002rank}%
  \BibitemOpen
  \bibfield  {author} {\bibinfo {author} {\bibfnamefont {S.}~\bibnamefont {Burer}}, \bibinfo {author} {\bibfnamefont {R.~D.}\ \bibnamefont {Monteiro}},\ and\ \bibinfo {author} {\bibfnamefont {Y.}~\bibnamefont {Zhang}},\ }\bibfield  {title} {\bibinfo {title} {Rank-two relaxation heuristics for max-cut and other binary quadratic programs},\ }\href@noop {} {\bibfield  {journal} {\bibinfo  {journal} {SIAM Journal on Optimization}\ }\textbf {\bibinfo {volume} {12}},\ \bibinfo {pages} {503} (\bibinfo {year} {2002})}\BibitemShut {NoStop}%
\bibitem [{\citenamefont {Kirkpatrick}\ \emph {et~al.}(1983)\citenamefont {Kirkpatrick}, \citenamefont {Gelatt},\ and\ \citenamefont {Vecchi}}]{1983_Kirkpatrick_Science}%
  \BibitemOpen
  \bibfield  {author} {\bibinfo {author} {\bibfnamefont {S.}~\bibnamefont {Kirkpatrick}}, \bibinfo {author} {\bibfnamefont {C.~D.}\ \bibnamefont {Gelatt}},\ and\ \bibinfo {author} {\bibfnamefont {M.~P.}\ \bibnamefont {Vecchi}},\ }\bibfield  {title} {\bibinfo {title} {{Optimization by Simulated Annealing}},\ }\href {https://doi.org/10.1126/science.220.4598.671} {\bibfield  {journal} {\bibinfo  {journal} {Science}\ }\textbf {\bibinfo {volume} {220}},\ \bibinfo {pages} {671} (\bibinfo {year} {1983})}\BibitemShut {NoStop}%
\bibitem [{\citenamefont {Honjo}\ \emph {et~al.}(2021)\citenamefont {Honjo}, \citenamefont {Sonobe}, \citenamefont {Inaba}, \citenamefont {Inagaki}, \citenamefont {Ikuta}, \citenamefont {Yamada}, \citenamefont {Kazama}, \citenamefont {Enbutsu}, \citenamefont {Umeki}, \citenamefont {Kasahara} \emph {et~al.}}]{honjo2021100}%
  \BibitemOpen
  \bibfield  {author} {\bibinfo {author} {\bibfnamefont {T.}~\bibnamefont {Honjo}}, \bibinfo {author} {\bibfnamefont {T.}~\bibnamefont {Sonobe}}, \bibinfo {author} {\bibfnamefont {K.}~\bibnamefont {Inaba}}, \bibinfo {author} {\bibfnamefont {T.}~\bibnamefont {Inagaki}}, \bibinfo {author} {\bibfnamefont {T.}~\bibnamefont {Ikuta}}, \bibinfo {author} {\bibfnamefont {Y.}~\bibnamefont {Yamada}}, \bibinfo {author} {\bibfnamefont {T.}~\bibnamefont {Kazama}}, \bibinfo {author} {\bibfnamefont {K.}~\bibnamefont {Enbutsu}}, \bibinfo {author} {\bibfnamefont {T.}~\bibnamefont {Umeki}}, \bibinfo {author} {\bibfnamefont {R.}~\bibnamefont {Kasahara}}, \emph {et~al.},\ }\bibfield  {title} {\bibinfo {title} {100,000-spin coherent ising machine},\ }\href {https://doi.org/10.1126/sciadv.abh0952} {\bibfield  {journal} {\bibinfo  {journal} {Science advances}\ }\textbf {\bibinfo {volume} {7}},\ \bibinfo {pages} {eabh0952} (\bibinfo {year} {2021})}\BibitemShut {NoStop}%
\end{thebibliography}%

\section{\label{sec:supp_material}\sm}

\begin{figure*}
\centering
\begin{tabular}{|c|c|c|}
\hline
\subf{\includegraphics[width=40mm]{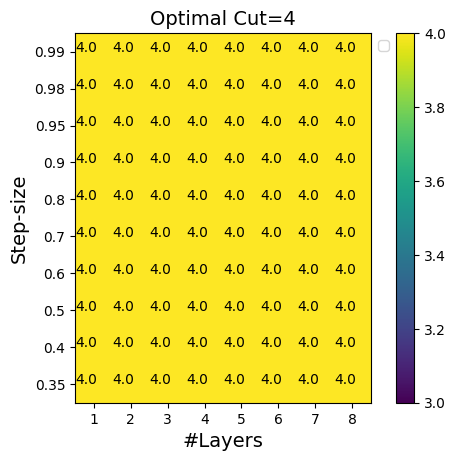}}
     {4 nodes}
&
\subf{\includegraphics[width=40mm]{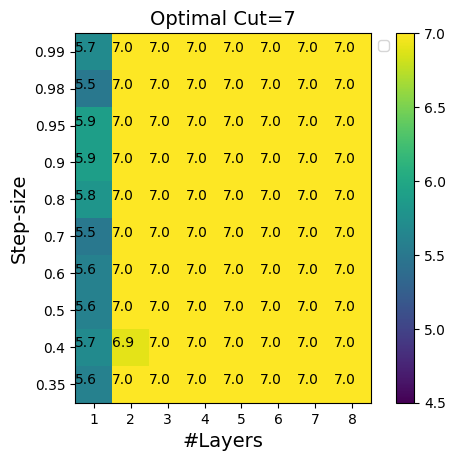}}
     {6 nodes}
&
\subf{\includegraphics[width=40mm]{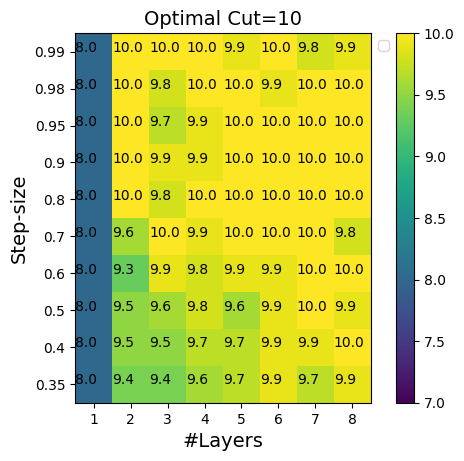}}
     {8 nodes}
\\
\hline
\subf{\includegraphics[width=40mm]{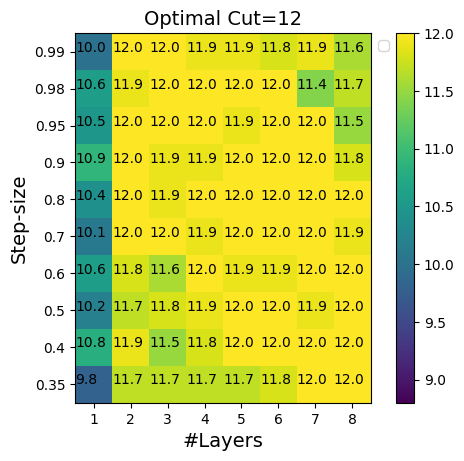}}
     {10 nodes}
&
\subf{\includegraphics[width=40mm]{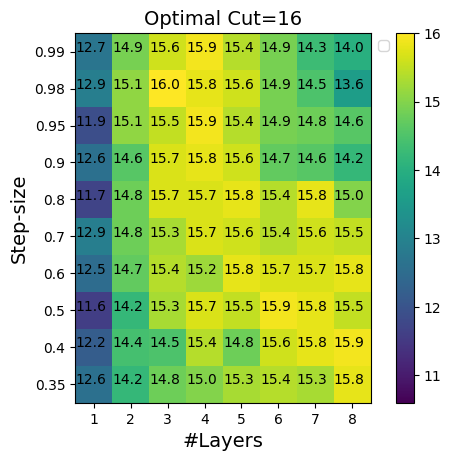}}
     {12 nodes}
&     
\subf{\includegraphics[width=40mm]{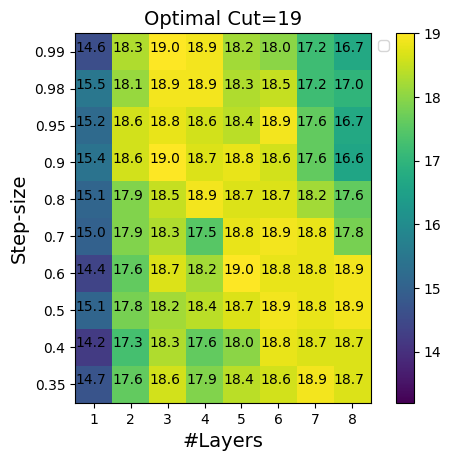}}
     {14 nodes}
\\
\hline
\subf{\includegraphics[width=40mm]{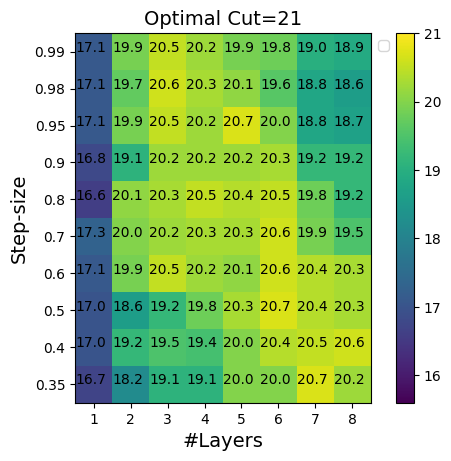}}
     {16 nodes}
&
\subf{\includegraphics[width=40mm]{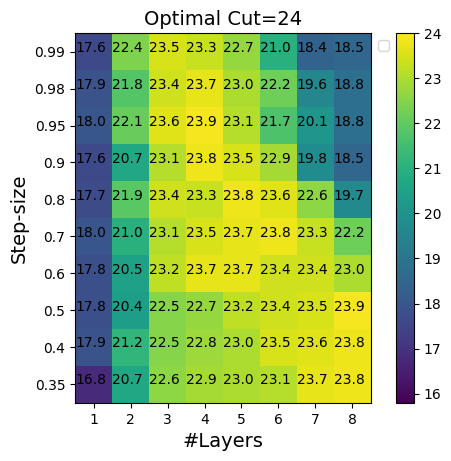}}
     {18 nodes}
&
\subf{\includegraphics[width=40mm]{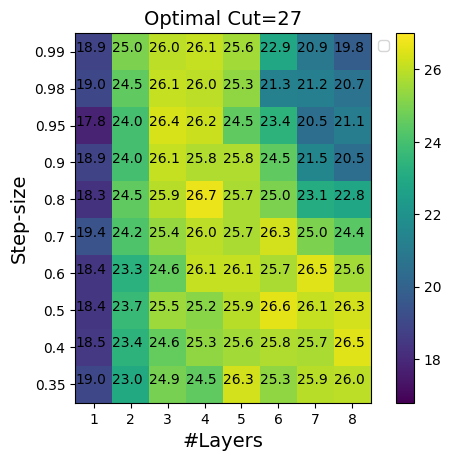}}
     {20 nodes}
\\
\hline
\subf{\includegraphics[width=40mm]{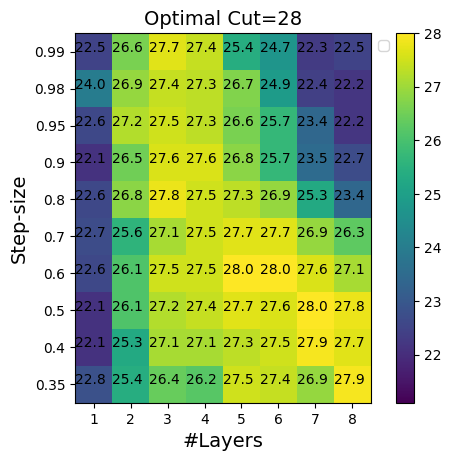}}
     {22 nodes}
&
\subf{\includegraphics[width=40mm]{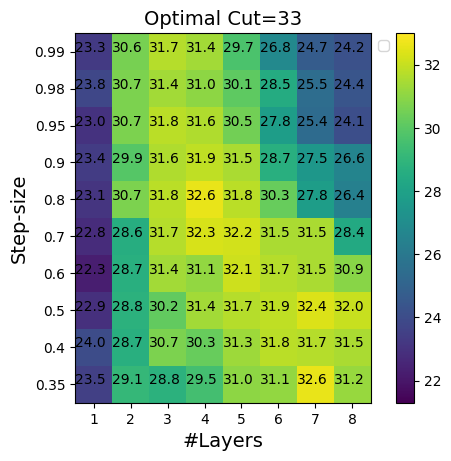}}
     {24 nodes}
&
\subf{\includegraphics[width=40mm]{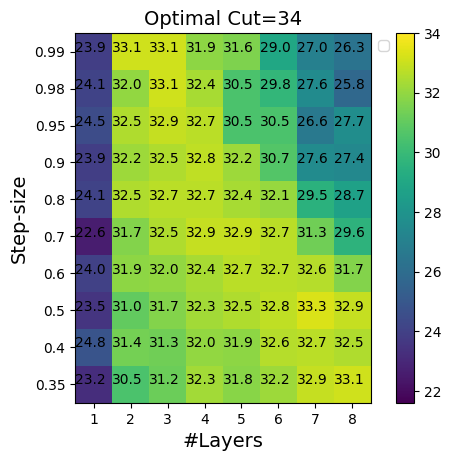}}
     {26 nodes}
\\
\hline
\subf{\includegraphics[width=40mm]{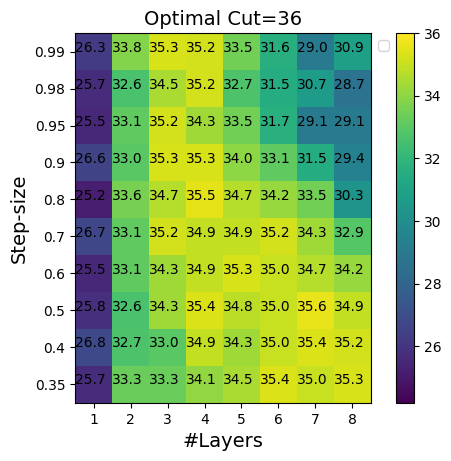}}
     {28 nodes}
&
\subf{\includegraphics[width=40mm]{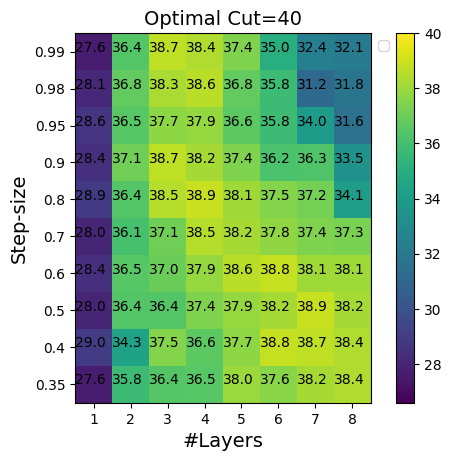}}
     {30 nodes}
&
\subf{\includegraphics[width=40mm]{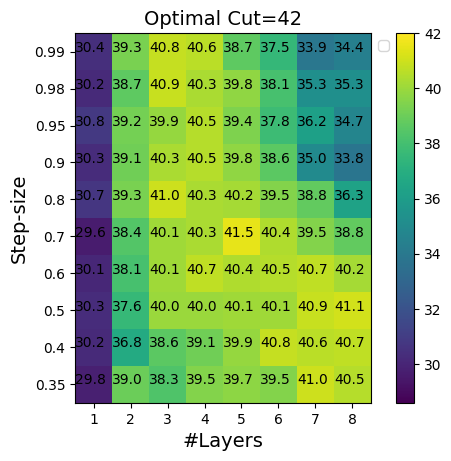}}
     {32 nodes}
\\
\hline
\end{tabular}
\caption{Grid-searches of the number of layers and step-size combinations for the set of 3-regular graphs with 4-32 nodes, with one graph instance per graph-size, described in Sec.~\ref{secsec:4-32-single}.  
A grid-search is presented for each graph instance. 
Each $\{$Layers, $\#$Step-size$\}$ entry displays the averaged QEMC cut obtained over 10 trials via noiseless state-vector simulations, where each trial consisted of 300 iterations. 
The optimal cut, obtained by exhaustive search, is also specified for comparison.}
\label{fig: grid_search_params}
\end{figure*}

\begin{figure*}[h!]
    \centering
    \includegraphics[width=\textwidth]{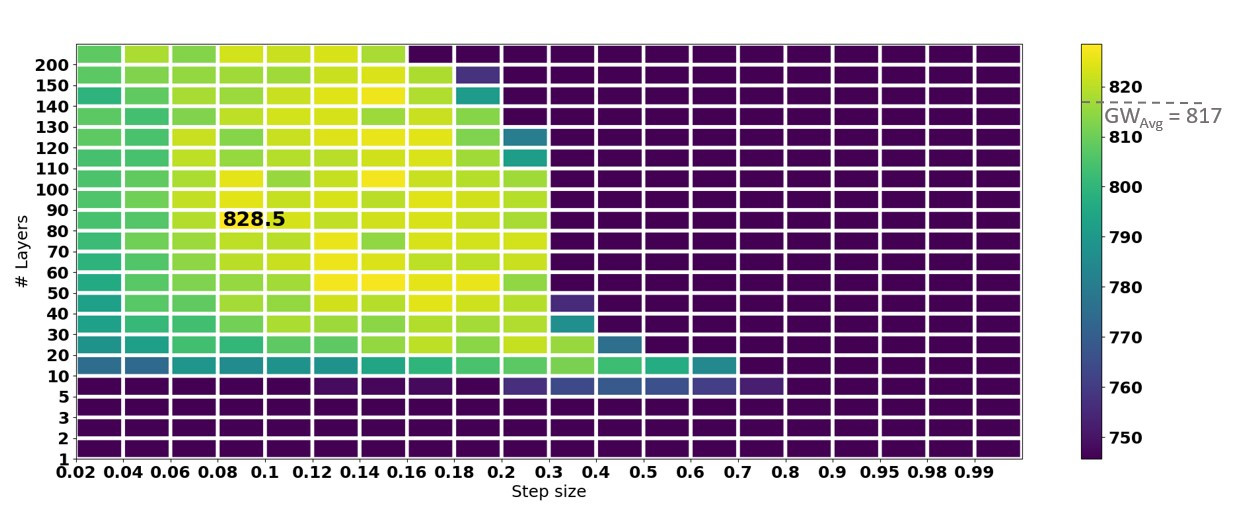}
    \caption{Grid-search of the number of layers and step-size combinations for the single 9-regular, 256-node graph instance, described in Sec.~\ref{secsec:16-2048-single}. Each $\{\#$Layers, Step-size$\}$ entry represents the averaged QEMC cut obtained over 10 trials via noiseless state-vector simulations, where each trial consisted of 200 iterations. The optimal hyperparameters combination is identified as $\#$layers = 80 and step-size=0.08, resulting in an average cut of 828.5. The averaged GW cut of 817 is also specified along the color bar for comparison.}
    \label{fig:grid_search_256}
\end{figure*}

\paragraph{}
Fig.~\ref{fig: grid_search_params} presents the hyperparameters grid-searches for the 3-regular graphs set, ranging from 4 to 32 nodes, with one graph instance per graph-size, as described in Sec.~\ref{secsec:4-32-single}. The search was performed for each graph instance. It is observed that utilizing more layers  often requires a smaller step-size of the Adam optimizer.
Similarly, Fig.~\ref{fig:grid_search_256} shows the parameters grid-search for the $256$-node graph instance of Sec.~\ref{secsec:16-2048-single}, illustrating the same trend. 

\begin{figure}[h!]
    \centering
    \includegraphics[width=0.48\textwidth]{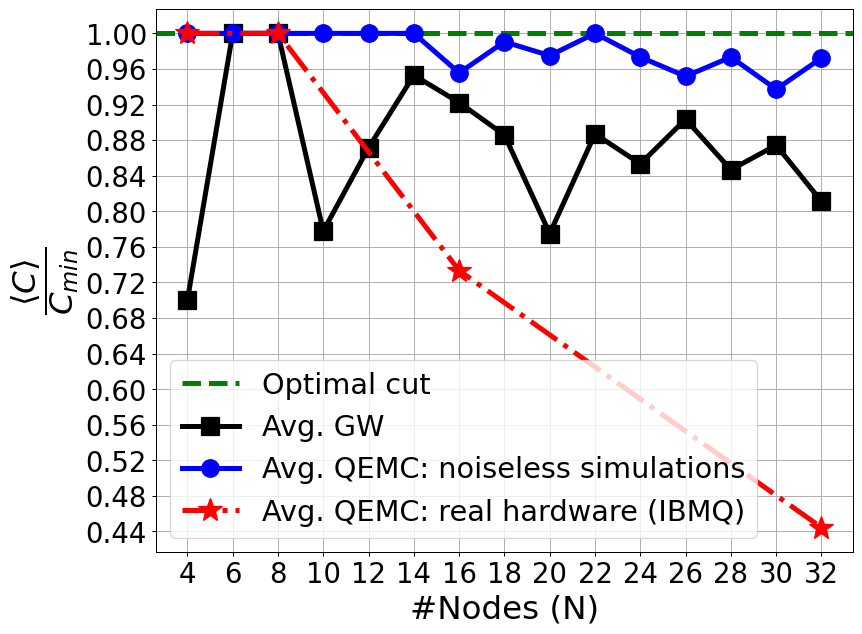}
    \caption{The exact same plot of Fig.~\ref{fig:best_res_4_32}, shown on the $y$-scale of \cite{harrigan2021quantum} for easier comparison.}
    \label{fig:google_scale}
\end{figure}

\paragraph{}
To facilitate a meaningful comparison between the results presented in Sec.~\ref{secsec:4-32-single} and those reported in Ref.~\cite{harrigan2021quantum}, Fig.~\ref{fig:google_scale} adjusts the $y$-axis scaling of Fig.~\ref{fig:best_res_4_32}. This rescaling is performed to align the metric used in Ref.~\cite{harrigan2021quantum}, denoted as $\frac{\langle C \rangle}{C_{\min}}$, with our metric $\frac{\text{Cut}}{\text{Cut*}}$, where 
$C = \sum_{j < k} Z_j Z_k$ is the cost function utilized in Ref.~\cite{harrigan2021quantum} (for unweighted graphs) and   
$C_{\min}$ is the minimal value it takes, while $\text{Cut}$ is the averaged (over multiple QEMC executions) \textit{cut} and $\text{Cut*}$ is the optimal \textit{cut}. Here, the quantity \textit{cut} accounts for the number of edges that connect nodes of different colors, as defined earlier. In contrast, according to the definition of $C$,  
edges that connect nodes of different colors contribute $-1$, whereas edges that connect nodes of the same color contribute $+1$ to $C$, resulting with $\langle C \rangle = -Cut + (M - Cut) = M - 2Cut$ and $C_{min} = -Cut^* + (M - Cut^*) = M - 2Cut^*$, leading to: 
\begin{eqnarray}
    \frac{\langle C \rangle}{C_{min}} = \frac{M - 2Cut}{M - 2Cut^*},
\end{eqnarray}
where $M = |E|$ is the number of edges in the graph.

\end{document}